\documentclass[lettersize,journal]{IEEEtran}
\usepackage{amsmath,amsfonts,amssymb}
\usepackage{algpseudocode}
\usepackage{algorithm}
\usepackage{array}
\usepackage[caption=false,font=normalsize,labelfont=sf,textfont=sf]{subfig}
\usepackage{bm}
\usepackage{textcomp}
\usepackage{xcolor}
\usepackage{makecell}
\usepackage{tabularx} 
\usepackage{threeparttable}
\usepackage{multirow} 
\usepackage{booktabs} 
\usepackage{stfloats}
\usepackage{url}
\usepackage{adjustbox}
\usepackage{verbatim}
\usepackage{graphicx}
\usepackage{framed}
\usepackage{cite}
\usepackage{arydshln}
\usepackage{tcolorbox}
\usepackage[colorlinks,
            linkcolor=blue,
            anchorcolor=black,
            citecolor=blue
            ]{hyperref}

\usepackage{tcolorbox}
\newtcolorbox{promptbox}{
  colback=blue!4!white,
  colframe=blue!65!black,
  boxrule=0.8pt,
  arc=3pt,
  left=6pt,
  right=6pt,
  top=6pt,
  bottom=6pt
}
\begin{document}

\title{TCLNet: A Hybrid Transformer–CNN Framework Leveraging Language Models as Lossless Compressors for CSI Feedback}

\author{Zijiu Yang, 
Qianqian Yang, 
Shunpu Tang, 
Tingting Yang, 
and
Zhiguo Shi

\thanks{This work was in part presented at the 42nd International Conference on Machine Learning (ICML) Workshop on Machine Learning for Wireless Communications (ML4Wireless), Vancouver, Canada, July 2025 \cite{ML4WC}.}
\thanks{Z. Yang,  Q. Yang, S. Tang and Z. Shi are all with the College of Information Science and Electronic Engineering, Zhejiang University, Hangzhou 310027, China (E-mails: \{zijiu\_yang,  qianqianyang20, tangshunpu, shizg\}@zju.edu.cn). }
\thanks{T. Yang is with Peng Cheng Laboratory, Shenzhen 518066, China (E-mail: yangtt@pcl.ac.cn).
}
\thanks{The corresponding author of this paper is Q. Yang.}
}

\IEEEpubid{}

\maketitle

\begin{abstract}
In frequency division duplexing (FDD) massive multiple-input multiple-output (MIMO) systems, downlink channel state information (CSI) plays a crucial role in achieving high spectrum and energy efficiency. However, the CSI feedback overhead becomes a major bottleneck as the number of antennas increases. Although existing deep learning–based CSI compression methods have shown great potential, they still face limitations in capturing both local and global features of CSI, thereby limiting achievable compression efficiency. To address these issues, we propose TCLNet, a unified CSI compression framework that integrates a hybrid Transformer–CNN architecture for lossy compression with a hybrid language-model (LM) and factorized-model (FM) design for lossless compression. The lossy module jointly exploits local features and global context, while the lossless module adaptively switches between context-aware coding and parallel coding to optimize the rate–distortion–complexity (RDC) trade-off. Extensive experiments on both real-world and simulated datasets demonstrate that the proposed TCLNet outperforms existing approaches in terms of reconstruction accuracy and transmission efficiency, achieving up to a 5 dB performance gain across diverse scenarios. Moreover, we show that large language models (LLMs) can be leveraged as zero-shot CSI lossless compressors via carefully designed prompts.

\end{abstract}

\begin{IEEEkeywords}
CSI feedback, hybrid  transformer-CNN architecture, language models
\end{IEEEkeywords}

\section{Introduction}
\subsection{Background}
\begin{figure}[t!]
	\begin{center}
		\centerline{\includegraphics[width=\columnwidth]{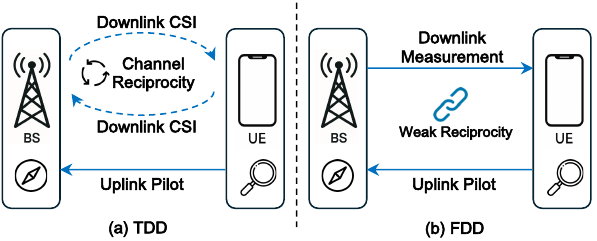}}
		\caption{Illustration of  CSI feedback in Massive MIMO systems. In FDD systems, the lack of reciprocity between uplink and downlink CSI makes direct estimation of downlink CSI challenging.}
		\label{graph1}
	\end{center}
\end{figure}

Massive multiple-input multiple-output (MIMO) is a key technology for 5G and is expected to play an even more significant role in future 6G systems. By deploying hundreds or even thousands of antennas at the base station, massive MIMO can greatly enhance spectrum and energy efficiency, while also enabling better spectral efficiency for a large number of users and devices \cite{I1}. To fully realize these benefits, the base station (BS) needs accurate downlink channel state information (CSI) to support precoding, resource allocation, and interference management \cite{I2}. As shown in \autoref{graph1}, in time division duplexing (TDD) systems, the base station exploits channel reciprocity and estimates downlink CSI directly from uplink measurements. In contrast, frequency division duplexing (FDD) systems use separate frequency bands for uplink and downlink. In this case, only weak reciprocity exists, making it very challenging to infer downlink CSI from uplink CSI \cite{I3}. Consequently, the user equipment (UE) must estimate the downlink CSI and then feed it back to the BS.

However, as the number of antennas continues to scale in massive MIMO systems, the CSI dimension grows large correspondingly, resulting in a prohibitive feedback burden. To address this issue, researchers have developed efficient CSI compression and feedback methods to reduce the overhead, ranging from classical approaches such as compressed sensing \cite{CS_1,CS_2} to more recent learning-based techniques \cite{Csinet,CRNet,CLNet,Transnet,CFnet,Csinet+,CsiPPP}.

\subsection{Related Works and Motivations}
In the early stage of MIMO research, traditional CSI compression methods, such as compressed sensing (CS), were widely explored in \cite{traditional_1,traditional_2,traditional_3,traditional_4}. However, their performance is limited in practice due to mismatches between idealized assumptions and practical channel conditions. Moreover, most CS algorithms depend on iterative optimization, which results in high computational complexity and non-negligible processing latency. To overcome these limitations, recent studies have explored learning-based solutions. Specifically, these approaches can be divided into two main categories \cite{L2O}:

\subsubsection{Model-driven methods} 
Model-driven methods follow the traditional CS theoretical framework but replace some steps with neural networks, which can maintain explainability while enhancing performance. A common strategy is to unfold iterative optimization algorithms into neural network architectures. For example, ISTA-Net \cite{ISTA} unfolds the iterative shrinkage–thresholding algorithm (ISTA), a classical solver for sparse recovery in CS, into a deep network, where the transformation matrices and thresholding functions are learned directly from data to improve CSI reconstruction.  Building on this idea, more advanced algorithms such as TiLISTA-Joint \cite{TiLISTA-Joint} and FISTA-Net \cite{FISTA-Net} enhance the unfolded CS framework by incorporating improved update rules, adaptive thresholds, and momentum-based acceleration techniques, leading to faster convergence and higher feedback efficiency.

Despite their effectiveness, model-driven strategies still face several challenges, as they truncate iterative algorithms into a fixed number of layers. This fixed-depth design prevents the model from adapting the number of iterations to different channel conditions. Since more iterations are needed when the channel is highly sparse or the compression ratio is low, while fewer iterations would suffice for easier cases \cite{L2O}. More importantly, this truncation weakens the theoretical convergence guarantees of the original optimization algorithm. To alleviate these issues, recent work such as Csi-L2O \cite{Csi-L2O} learns the optimization rule directly instead of unfolding a specific algorithm. By using a linear projection at the user side and an LSTM-based learned optimizer at the BS, it supports variable compression ratios and allows the update rule to generalize across different problem sizes. However, Csi-L2O still relies on step-by-step iterative updates, which may converge slowly and become less robust under fast-varying channel conditions.

\subsubsection{Data-driven methods}
In contrast to model-driven approaches, data-driven methods directly learn end-to-end CSI compression and reconstruction from large CSI datasets. In this direction, the authors in \cite{Csinet} proposed the first end-to-end learning-based CSI compression framework, namely CSINet, demonstrating that CNN-based autoencoder can achieve significant gains in CSI feedback compared to the traditional CS approaches. Since then, a wide range of extensions has been proposed. For example, to better model spatial diversity, multi-resolution techniques were introduced in \cite{Csinet+,CRNet,DCRNet,MAMR,FOCU}, showing that large receptive fields are important to capture multi-scale propagation characteristics. Moreover, ImCSiNet \cite{Imcsinet}, and EVCsiNet \cite{Evcsinet} highlight the importance of leveraging advanced backbone networks to extract more expressive features. To improve the robustness, adaptive and domain-specific designs were introduced in \cite{Adalicsinet, DNNet, CFnet} to handle varying channel conditions. However, these methods still struggle to capture long-range dependencies in CSI. Although attention mechanisms can be used to enhance global context aggregation \cite{Atten-csi,CS-ReNet,DFECsiNet}, their effectiveness remains constrained by the underlying CNN architecture.

Recognizing this limitation, recent works have begun exploring transformer-based architectures \cite{Transformer}, since their self-attention mechanism can effectively capture long-range correlations. Specifically, the authors in \cite{Transnet} propose a dual-layer transformer module to extract informative global CSI features. Nevertheless, pure transformer architectures often incur higher computational and memory overhead compared with CNN-based designs, which limits their practical deployment. To bridge this gap, a natural direction is to integrate CNN and transformer architectures within a unified framework. However, the straightforward sequential CNN-transformer design in \cite{hybrid_tc} limits the potential benefits of such a hybrid architecture. In contrast, parallel integration might offer better complexity–performance trade-offs, as shown in recent work on image compression \cite{TCM}. \textit{Motivated by this, we aim to design a parallel hybrid Transformer–CNN architecture for CSI compression that balances global modeling capability and computational efficiency.} 
In addition to this, we aim to further encode the obtained low-dimensional representation of CSI using lossless compressors to reduce feedback overhead. Most learning-based methods do not explicitly take this coding process into account or simply perform fixed-precision scalar or vector quantization on the compressed features \cite{Csinet+, Csi-L2O, VQ, VQ-VAE}, which limits their achievable compression efficiency.

Therefore, introducing variable-length coding becomes important, as it can allocate bits according to the statistical structure of the feature distribution, thereby improving compression efficiency. However, its effectiveness depends on accurately estimating the probability of each symbol. Prior works such as \cite{DeepCMC} used factorized models (FMs) to estimate such probabilities, but FMs assume statistical independence among symbols and ignore contextual relationships within CSI, leading to suboptimal coding efficiency. \textit{Therefore, a probability model that can capture the complex dependencies in CSI is key to enabling more effective variable-length coding.}

With the recent successes of language models in modeling complex probability distributions, there is growing interest in using them as powerful probabilistic estimators for data compression. For example, LMC \cite{LMC} shows that trained language models (LMs) can act as general compressors by flattening non-text data into symbol sequences and predicting their probabilities for entropy coding. The authors in \cite{LLM4Image} further explore the use of large language models (LLMs) for image compression by introducing visual prompts that guide the model to capture image residual structures more effectively, thereby enabling more accurate symbol probability estimation and improved compression efficiency. Besides, in closely related wireless communication tasks, recent studies have demonstrated the potential of LMs in tasks such as channel prediction \cite{LLM4CP, li2025bridging, SCA_LLM} and channel estimation \cite{LLM4MT,WirelessGPT}, showing strong performance and generalization across scenarios. Although these works do not explicitly consider CSI coding or compression, they indicate that LMs can capture and understand channel characteristics.
Additionally, several modern smartphones are equipped with built-in on-device language models, such as Vivo BlueLM\footnote{\url{https://github.com/vivo-ai-lab/BlueLM}} and Apple Intelligence\footnote{\url{https://machinelearning.apple.com/research/apple-intelligence-foundation-language-models}} on the iPhone. These embedded language models can be further exploited to support and enhance physical-layer communication tasks.
\textit{Motivated by these, we aim to leverage LMs to build more expressive probability models for lossless CSI coding.}

\subsection{Contributions}
In this paper, we investigate CSI feedback in the FDD massive MIMO-OFDM system. We propose a novel CSI compression framework named TCLNet, which consists of a lossy compressor and a lossless compressor. Specifically, for the lossy compressor, we integrate CNN and Transformer architectures in a parallel hybrid design to capture long-range dependencies and fine-grained local features. To this end, we design the lossy encoder and decoder based on a specialized feature extraction block, termed the \emph{TransConv} module, which splits the input features into local and global pathways.  The local pathway uses asymmetric convolutions to capture fine-grained details, while the global pathway adopts Swin Transformers to model long-range dependencies. By fusing these features through residual connections, the architecture reduces the computational overhead of pure Transformers and achieves better reconstruction than conventional CNNs.

Following the lossy compression, we propose a variable-complexity lossless compressor based on language modeling to further reduce the feedback overhead. Specifically, we first leverage a suitably scaled context-aware LM as a distribution predictor for CSI symbols, which improves entropy-coding efficiency but introduces high computational latency due to its autoregressive nature. To mitigate this issue, we introduce a lightweight factorized model (FM) \cite{minnen,balle} together with a novel symbol selection mechanism, which dynamically allocates computational resources by processing information-dense symbols through the autoregressive LM and statistically independent symbols through the parallel FM, thereby explicitly addressing the rate–distortion–complexity (RDC) trade-off. 
Additionally,
we further explore the potential of LLMs as zero-shot CSI compressors, demonstrating that LLMs are capable of predicting CSI symbols with designed prompts.

The main contributions of this paper can be summarized as follows
\begin{itemize}
   \item We propose a parallel hybrid Transformer–CNN lossy compressor that fully exploits the complementary strengths of both Transformer and CNN, effectively balancing computational complexity and CSI reconstruction performance.

    \item We introduce a hybrid LM–FM lossless compression framework that dynamically routes symbols to either an autoregressive LM or a parallel FM based on a novel symbol-selection mechanism. This enables fine-grained control over computational complexity and achieves highly efficient variable-length coding. Furthermore, we explore the integration of LLMs through prompt engineering for lossless CSI compression.

    \item We conduct extensive experiments on both real-world and simulated datasets to demonstrate that the proposed framework can outperform existing CSI feedback methods in terms of reconstruction performance and compression efficiency. Moreover, we show that LLMs can function as zero-shot lossless CSI compressors, without any domain-specific training.
\end{itemize}

\section{System Model}
\label{sec:system_model}
In this section, we introduce the system model of this paper. Specifically, we first present the channel model and then introduce the process of CSI feedback.
\subsection{Channel Model}
We consider a typical FDD massive MIMO-OFDM system, where there are $N_t \gg 1$
transmit antennas at the BS, and each UE is equipped with a single receive antenna. The system bandwidth is divided into $N_c$ subcarriers in the frequency domain. Let $\bm{h}_n \in \mathbb{C}^{N_t \times 1}$ denote the downlink channel gain vector, and the received signal at the $n$-th sub-carrier can be given by
\begin{align}\label{receive signal}
    y_n=\bm{h}_{n}^H\bm{w}_n x_n+\epsilon_n,
\end{align}
where $x_n \in \mathbb{C}$ denotes the transmitted data symbol, $\bm{w}_n$ is the precoding vector and $\epsilon_n \sim \mathcal{CN}(0,\sigma^2)$ is additive white Gaussian noise.  

By stacking all subcarriers in the spatial frequency domain, the overall CSI matrix can be written as
\begin{equation}
	\bm{H} = 
	\begin{bmatrix}
		h_{0,0} & h_{0,1} & ... &h_{0,N_t-1} \\
		h_{1,0} & h_{1,1}& ... &h_{1,N_t-1}\\
		\vdots & \vdots & \ddots & \vdots \\
		h_{N_c-1,0} & h_{N_c-1,1}& ... &h_{N_c-1,N_t-1}\\
	\end{bmatrix}
\end{equation}
where $\bm{H}\in\mathbb{C}^{N_c\times N_t}$ denotes the CSI matrix, in which the $i$-th row corresponds to the channel coefficients of the $i$-th subcarrier, and the $j$-th column represents the channel response associated with the $j$-th transmit antenna. In FDD systems, such a CSI matrix must be fed back to the BS, thereby optimizing the precoding vector $\bm{w}_n$ for each subcarrier to improve the received signal quality and the system spectral efficiency. However, since the CSI matrix has a real dimension of $2N_cN_t$, the feedback overhead significantly increases with the number of subcarriers $N_c$ and transmit antennas $N_t$. Such a large amount of feedback not only costs significant uplink transmission resources but also becomes impractical in future massive MIMO-OFDM systems. Therefore, it is necessary to compress the CSI matrix before feedback.

\subsection{CSI Feedback}

The CSI feedback procedure typically involves three main stages: compression, quantization, and reconstruction. Specifically, following \cite{Csinet}, we first transform the CSI from the spatial-frequency domain to the angle-delay domain using a 2D discrete Fourier transform (DFT), given by
\begin{equation}
 \bm{H}'=\bm{F}_c \bm{H} \bm{F}_t ,
\end{equation}
where $\bm{H}’$ denotes the transformed CSI matrix in the angle–delay domain. Besides, $\bm{F}_c \in \mathbb{C}^{N_c \times N_c}$ and $\bm{F}_t \in \mathbb{C}^{N_t \times N_t}$ denote the DFT transformation matrices along the frequency and spatial dimensions, respectively. This transformation concentrates most of the channel energy into a few dominant coefficients located in the first $N_a$ rows $(N_a \leq N_c)$, while the remaining entries are nearly zero. Thus, $\bm{H}'$ can be approximated by retaining only its first $N_a$  rows, resulting in a lower-dimensional representative matrix $\bm{H}_a \in \mathbb{C}^{N_a \times N_t}$. Next, to further reduce the feedback overhead, the low-dimensional matrix $\bm{H}_a$ is compressed and quantized using a learning-based autoencoder framework.  Specifically, we use $E_l(\cdot)$ and $E_{nl}(\cdot)$ to denote the lossy encoder and the lossless encoder at the UE, respectively, and we employ $Q(\cdot)$ and $Q_d(\cdot)$ to denote the quantizer and dequantizer. The compression with a predefined ratio $CR$ and the quantization process can be expressed as follows
\begin{align}
\label{e1}
\bm{z} &= E_l(\bm{H}_a),\\
\bm{z'} &= Q(\bm{z}),\\
\bm{b} &= E_{nl}(\bm{z'}),
\end{align}
where $\bm{z}\in \mathbb{R}^{z} $ denotes the compressed CSI feature with $z = N_a \times N_t \times \text{CR}$,
$\bm{z}^\prime \in \mathbb{R}^{Z}$ denotes the quantized symbols and $\bm{b}$ represents the quantized bitstream for feedback.

At the receiver, a lossy decoder $D_l(\cdot)$ and a lossless decoder $D_{nl}(\cdot)$ are employed to reconstruct the CSI from the received $\bm{b}$, given by
\begin{align}
		\bm{z'} = D_{nl}(\bm{b}),\\
        \hat{\bm{z}} = {Q}^{-1}(\bm{z'}),\\
		\hat{\bm{H}}_a = D_{l}( \hat{\bm{z}} ),
\end{align}
where $\bm{z'}$ and $\bm{H}_a'$ represent the decoded vector and the reconstructed CSI, respectively.

With this architecture, the key challenge is how to design $E_l(\cdot), E_{nl}(\cdot)$ and $D_l(\cdot), D_{nl}(\cdot)$ to achieve efficient CSI compression and reconstruction. However, most prior works \cite{Csinet+, Transnet, CRNet, CLNet, DCRNet} only focused on fixed-length coding, improving NMSE under a fixed bit rate while neglecting rate control. Therefore, it is necessary to jointly consider bit rate, distortion, and computational complexity to address the RDC trade-off in CSI feedback.

\begin{figure*}[!t]
	\centering
	\centerline{\includegraphics[width=0.85\textwidth]{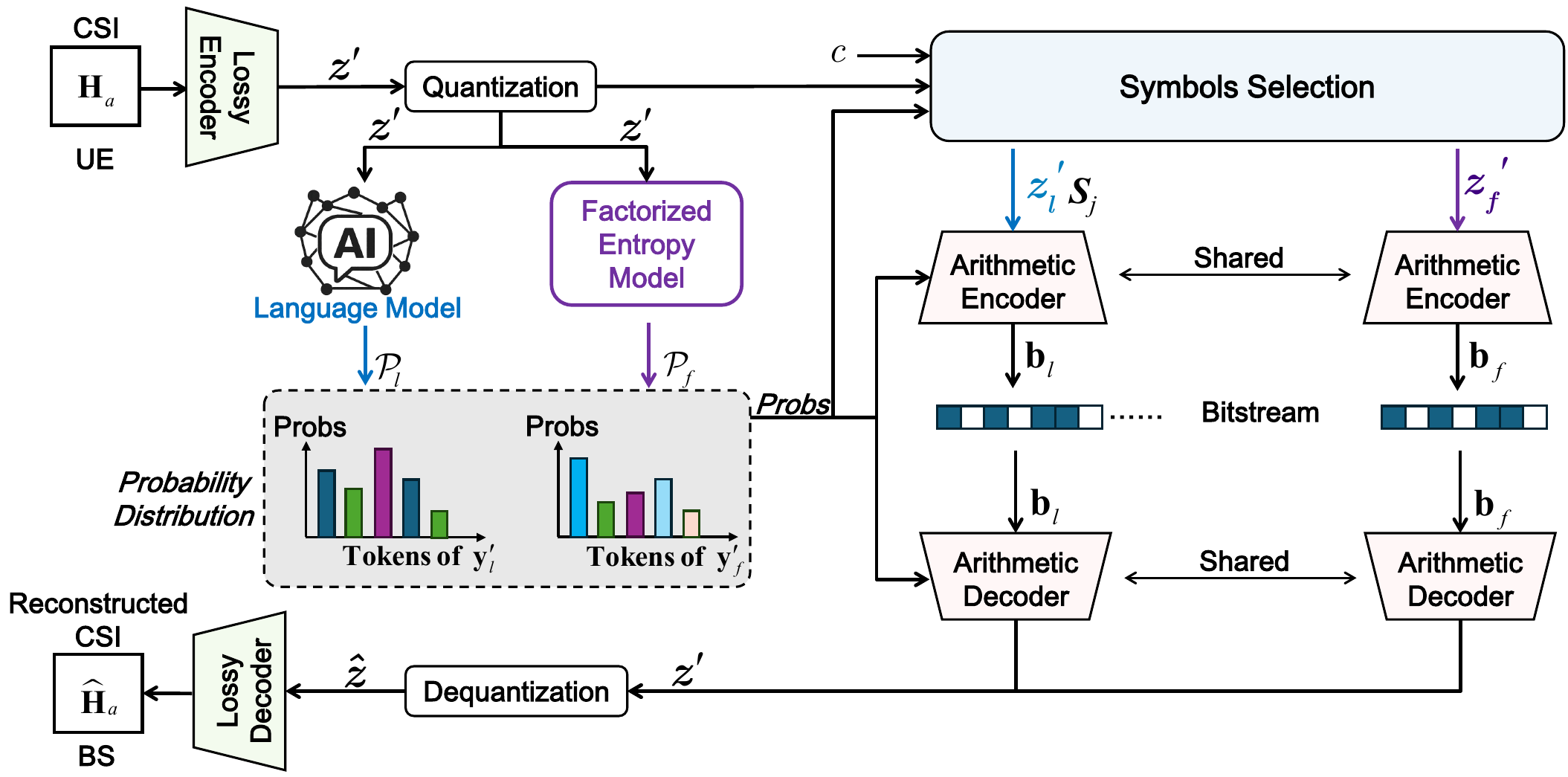}}
	\caption{Illustration of the proposed TCLNet, which consists of a lossy encoder based on a Transformer-CNN hybrid architecture and a lossless encoder based on LM and FM. The lossy encoder is used to compress the amount of CSI, while the lossless encoder is used for estimating the information entropy and encoding.
}
	\label{graph3}
\end{figure*}
\section{Proposed TCLNet}
\label{sec:TCLNet}
In this section, we introduce the proposed TCLNet for efficient CSI feedback. Specifically, we first present an overview of TCLNet, followed by detailed descriptions of its lossy and lossless compressors. Next, we discuss the training strategy for TCLNet and discuss the complexity of the proposed method.

\subsection{Overview}
As illustrated in \autoref{graph3}, the proposed TCLNet consists of a hybrid  Transformer–CNN-based lossy compressor and a hybrid LM–FM-based lossless compressor. The lossy compressor, consisting of a lossy encoder and a lossy decoder, is designed to extract the most informative features from the high-dimensional CSI, whereas the lossless compressor further compresses the quantized latent features into a bitstream for transmission. Specifically, at the transmitter, the lossy encoder first encodes the CSI $\bm{H}_a$ into a compact representation $\bm{z}$, which is then quantized into discrete symbols $\bm{z}'$.  These symbols are then input to the lossless compressor, which includes two parallel branches implemented using LM and FM modules and an arithmetic codec.

The LM and FM, denoting language models and factorized models respectively, generate the probability distributions $\mathcal{P}_l$ and $\mathcal{P}_f$ for entropy coding. Subsequently, these distributions ($\mathcal{P}_l, \mathcal{P}_f$), the complexity variable $c$, and the quantized features $\bm{z}'$ are input into the symbol selection module. This module generates the selection indicator $\bm{S}_j$. Specifically, $\bm{S}_j$ indicates the subset of elements in $\bm{z}'$  which are encoded by the Arithmetic Encoder using $\mathcal{P}_l$ to yield the bitstream $\bm{b}_l$, while the rest of $\bm{z}'$  are encoded using $\mathcal{P}_f$ to yield $\bm{b}_f$.
At the receiver, the Arithmetic Decoder executes the decoding process. It autoregressively decodes the bitstream $\bm{b}_l$ conditioned on the LM, and simultaneously decodes $\bm{b}_f$ conditioned on the FM, thereby recovering the quantized symbols $\bm{z}'$. Finally, the recovered $\bm{z}'$ is dequantized and passed to the lossy decoder to obtain the reconstructed CSI $\hat{\bm{H}}_a$.

\begin{figure*}[!t]
	\vskip 0.2in
	\centering
	\centerline{\includegraphics[width=0.9\textwidth]{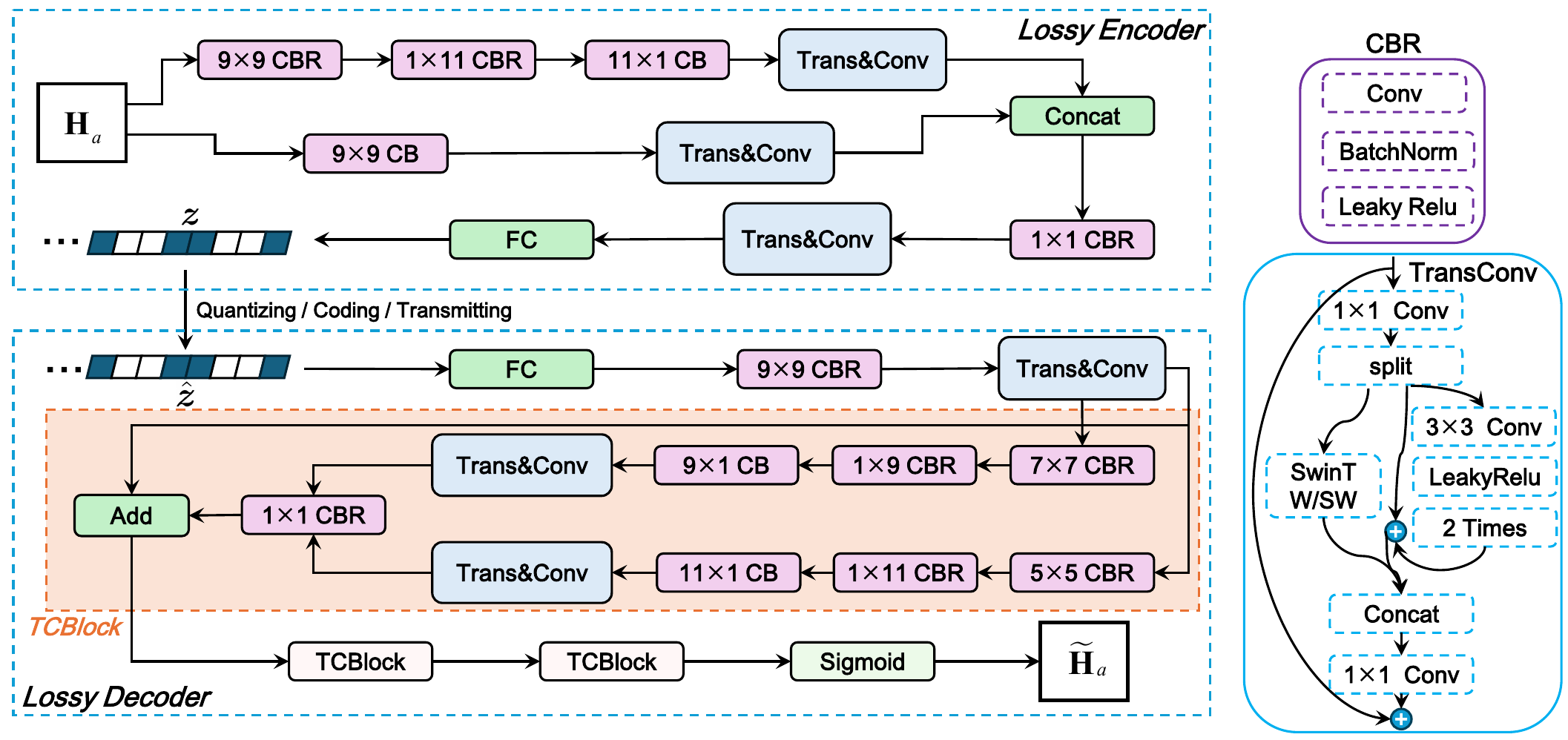}}
	\caption{Illustration of the proposed hybrid  transformer-CNN-based lossy compressor, which is composed of both CNN and swin-transformer, focusing on extracting local and global features from the CSI, respectively.}
	\label{graph4}
	\vskip -0.2in
\end{figure*}
\subsection{Hybrid  Transformer-CNN-based Lossy Compressor}

To effectively address the challenge of capturing both local spatial features and long-range dependencies in CSI matrices, as encountered in existing works \cite{CLNet, Transnet}, we propose a novel hybrid  Transformer–CNN-based lossy compressor that leverages the respective strengths of CNNs and Transformers, as illustrated in \autoref{graph4}. The proposed lossy compressor comprises two main components: a lossy encoder deployed at UE and a lossy decoder at the BS.

\subsubsection{Encoder Architecture} At the UE side, the CSI matrix $\bm{H}_a$ is first fed into two parallel feature extraction branches, each designed to capture complementary aspects of the CSI information. This dual-branch design enables a multi-scale representation and enhances the overall feature extraction capability \cite{CRNet,DCRNet}. In the first branch, three convolutional blocks with batch normalization and ReLU (CBR) operations are applied sequentially to progressively enlarge the receptive field and capture spatial correlations at different scales. The first CBR block employs a standard $9 \times 9$ kernel, followed by two asymmetric CBRs with kernel sizes of $1 \times 11$ and $11 \times 1$, respectively. This is because asymmetric kernels serve as an efficient alternative to separable convolutions, effectively capturing spatial dependencies while maintaining computational efficiency. The output from these CBRs is then fed into a \emph{TransConv} module, a hybrid Transformer–CNN block designed to capture both global and local features. The detailed structure of the \emph{TransConv} module will be described later. The second branch is responsible for contextual feature extraction, starting with a $9 \times 9$ CBR followed by another \emph{TransConv} module. The outputs from both branches are concatenated and fused through a $1 \times 1$ CBR for channel-wise integration, followed by a final \emph{TransConv} module for further feature refinement. Finally, a fully connected (FC) layer compresses the refined feature map into a compact feature vector sequence $\bm{z}$ according to the desired compression ratio $\text{CR}$. These feature vectors are then quantized and transmitted.

\subsubsection{Decoder Architecture} At the BS side,  the de-quantized feature vector is first projected back to its original size using an FC layer, followed by a 9$\times$9 CBR and a TransConv module to integrate both local and global information. Moreover, we design three sequential \emph{TCBlocks} to progressively reconstruct the CSI matrix. Similar to the encoder, each TCBlock contains two parallel branches. The first branch applies $7 \times 7$, $1 \times 9$, and $9 \times 1$ convolutions in sequence, followed by a \emph{TransConv} module, while the second branch adopts $5 \times 5$, $1 \times 11$, and $11 \times 1$ CBR blocks followed by another \emph{TransConv} module. The outputs from the two branches are concatenated and passed through a $1 \times 1$ convolutional block for feature fusion. Inspired by residual learning \cite{ResNet}, a skip connection is added from the input to the output of each TCBlock to facilitate information flow and improve training stability. After passing through all three TCBlocks, the reconstructed CSI matrix can be obtained via a sigmoid activation function to ensure that the reconstructed values fall within the range of the original CSI.

\subsubsection{Details of the TransConv Module} 
As the core component of the proposed TCLNet, the TransConv module plays a crucial role in effectively capturing both local and global features from the CSI data. 
Motivated by parallel TransConv employment in image compression \cite{TCM}, the architecture of the TransConv module is further introduced for CSI compression, as shown in the right part of \autoref{graph4}. Specifically, let the input of the TransConv module be denoted as $\bm{F}\in \mathbb{R}^{2 \times h\times w}$ where the two channels come from the real and imaginary parts of the CSI feature map. We first perform feature channel expansion using a 1$\times$1 convolution to increase the number of feature channels from 2 to $r (r > 2)$, which enhances the feature representation capability, given by
\begin{equation}
	\begin{aligned}
		\bm{F}^\prime \in \mathbb{R}^{r \times h\times w}= \text{Conv}_{1 \times 1}(\bm{F}),
	\end{aligned}
\end{equation}
where $\text{Conv}_{1 \times 1}(\cdot)$ is the channel expansion operation and $\bm{F}^\prime$ denotes the expanded feature map. Then the feature representation undergoes a channel-wise split operation, where the channels are almost equally divided to enable both global and local feature extraction, which can be expressed as
\begin{equation}
	\begin{aligned}
		\bm{F}_g, \bm{F}_l = \text{Split}(\bm{F}^\prime),
	\end{aligned}
\end{equation}
where $\bm{F}_l\in \mathbb{R}^{r_l \times h\times w}, \bm{F}_g\in \mathbb{R}^{r_g \times h\times w}$ represent the features directed to the local and global pathways, respectively.
The local branch $\bm{F}_l$ is processed through two successive convolutions with small kernels, followed by a residual connection with the original input to obtain locally enhanced features $\bm{F}^\prime_l$. 
In parallel, the global branch $\bm{F}_g$ is passed through a Swin Transformer module to capture long-range dependencies and global context, obtaining global representations $\bm{F}^\prime_g$. The outputs $\bm{F}_{\text{o}} \in \mathbb{R}^{r \times h\times w}$ from both branches are then concatenated and projected via a 1$\times$1 convolution, fused with the original input through another residual connection, given by
\begin{equation}
\begin{aligned}
	\bm{F}_{\text{o}} = \text{Conv}_{1 \times 1}(\text{concat}(\bm{F}^\prime_l,\bm{F}^\prime_g)) + \bm{F}.
\end{aligned}
\end{equation}
\subsection{LM- and FM-based Lossless Compressor}\label{LFC}
\begin{figure*}[!t]

	\centering
    \vskip 0.2in
	\centerline{\includegraphics[width=\textwidth]{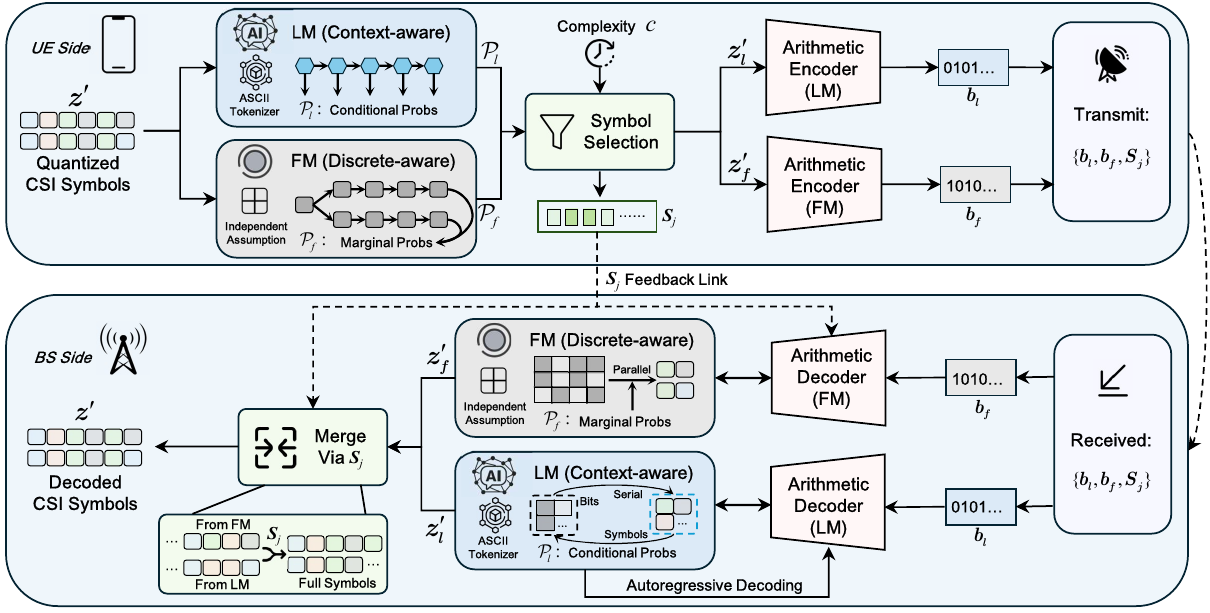}}
	\caption{Illustration of the proposed LM- and FM-based coding scheme, where the LM encodes latent symbols with statistical dependencies, while the FM encodes symbols that are relatively independent.}
	\label{graph5}
    \vskip -0.2in
\end{figure*}

After compressing the CSI into a compact feature vector via the lossy encoder, the representation is first quantized and then passed to a lossless compressor that produces the final bitstream for transmission. As illustrated in \autoref{graph5}, the lossless compressor processes the quantized symbols through two probability models and then applies symbol selection before performing arithmetic coding. In the following, we provide detailed descriptions of each component of the lossless compressor.

\subsubsection{Quantization and Probability Models}

The CSI feature vector $\bm{z}$ generated by the lossy encoder is first processed through an $n$-bit uniform quantizer,
given by
\begin{equation}
	\label{uniform_q}
	\begin{aligned}
		\Delta = \frac{\bm{z}_{max} - \bm{z}_{min}}{2^n - 1},\\
		\bm{z}^{\prime} =Q\left( \bm{z} \right) = \left\lfloor \frac{\bm{z}-\bm{z}_{min}}{\Delta} \right\rfloor,
	\end{aligned}
\end{equation}
where $\Delta$ denotes the quantization step size, $\bm{z}_{max}$ and $\bm{z}_{min}$ represent the maximum and minimum values of $\bm{z}$, respectively, and $\bm{z}^\prime$ denotes the quantized feature vector consisting of $2^n$ kinds of discrete symbols with sequence length of $z$.

The symbols are then passed to the LM and FM in parallel, obtaining the corresponding context-based probability distribution $\mathcal{P}_l$ and discrete probability distribution $\mathcal{P}_f$.
Specifically, the LM, denoted by $\mathrm{LM}(\cdot)$, first converts the discrete symbols into an ASCII token sequence and then predicts their context-dependent probabilities, given by

\begin{equation}
	\mathcal{P}_l  = \text{LM}(\bm{u}) = \left\{ \bm{\rho}_{0|d}, \bm{\rho}_{1|0,d}, \ldots, \bm{\rho}_{z \mid < z} \right\},
\end{equation}
where $\bm{u} = \text{ASCII}(\bm{z}^\prime)$ is the token sequence obtained from the quantized symbols $\bm{z}^\prime$ through ASCII tokenization. 
Besides, the notation $\bm{\rho}_{z\,|\,<z} \in \mathbb{R}^{|\mathbf{A}|}$ represents the context-based probability distribution for the $z$-th symbol, where the $z$-th token is predicted conditioned on all its preceding $z\!-\!1$ tokens. Here, $\mathbf{A}$ denotes the vocabulary of the LM, satisfying $|\mathbf{A}| > 2^n$, and $d$ denotes a dummy token used for sequence initialization or padding.

Moreover, we also employ a lightweight FM to estimate a context-free discrete probability distribution for the quantized symbols. The FM assumes statistical independence among symbols and models only their marginal distribution. Accordingly, the distribution $\mathcal{P}_f$ produced by the FM is given by
\begin{equation}
	\mathcal{P}_f = \mathrm{FM}(\mathbf{z}')
	= \left\{ \pi_{0}, \pi_{1}, \ldots, \pi_{2^n-1} \right\},
\end{equation}
where $\pi_i$ denotes the predicted marginal probability of the $i$-th symbol category and  $\sum_{i=0}^{2^n-1}\pi_i = 1$.

\subsubsection{Symbol Selection}
Next, we introduce the proposed symbol selection mechanism, which aims to adaptively select symbols for context-based compression using the LM, while assigning the remaining symbols to the FM for parallel processing. Specifically, we first compute the context-based entropy according to $\mathcal{P}_l$ for each token in $\bm{z}^\prime$ and the information entropy according to $\mathcal{P}_f$ for the entire symbol sequence, given by
\begin{align}
		H_{l,k}(\bm{z}^\prime_{k} \mid \bm{z}^\prime_{<k}) &=  - \sum_{a \in \mathbf{A}}  \rho_{k|<k}(a)\, \log_2 \rho_{k|<k}(a),\\
		H_f(\bm{z}^\prime)  & = - \sum_{s=0}^{2^n-1} \pi_s \log_2 \pi_s,
\end{align}
where $H_{l,k}(\bm{z}^\prime_{k} \mid \bm{z}^\prime_{<k})$ denotes the context-based entropy for the $k$-th token, and $H_f(\bm{z}^\prime)$ represents the discrete entropy for the entire symbol sequence, and $\rho _{k|<k}(a)$ is the context-based probability of token $k$ taking the vocabulary entry $a$.
Then, we can compute the entropy difference, given by
\begin{equation}
	\begin{aligned}
		D_k = H_f(\bm{z}^\prime) - H_{l,k}(\bm{z}^\prime_{i}|\bm{z}^\prime_{<i}),
	\end{aligned}
\end{equation}
where $D_k$ denotes the entropy difference for the $k$-th token. 

Based on the computed entropy differences, we can determine which symbols should be encoded using the LM or FM. To this end, we introduce a complexity variable $c \in (0,1)$ that specifies the proportion of symbols assigned to the FM for parallel processing. All tokens are first sorted according to their entropy differences. The top $(1-c)$ fraction with the largest differences are selected as $\bm{z}'_l$, and we use $\bm{S}_j$ to denote the corresponding indicator of the selected symbols. The remaining $c$ fraction of tokens with smaller entropy differences are grouped into $\bm{z}'_f$. This is because if a token satisfies
\begin{equation}
H_{l,k}(\bm{z}’_{k}|\bm{z}’_{<k}) \approx H_f(\bm{z}’),
\end{equation}
its context-dependent entropy provides little improvement over the context-free entropy, meaning contextual modeling contributes minimal compression gain. Such tokens are thus assigned to the FM branch for faster decoding. In contrast, when
\begin{equation}
H_{l,k}(\bm{z}’_{k}|\bm{z}’_{<k}) \ll H_f(\bm{z}’),
\end{equation}
the LM offers significantly better probability estimation due to strong contextual dependencies, and these tokens are thereby encoded using the LM.

\subsubsection{Arithmetic Codec and Dequantization}
After obtaining the probability distributions and the selected symbol sets, a statistical coding stage is applied to compress the quantized symbols. Specifically, we employ a finite-precision arithmetic coder \cite{Arithmetic_coding,LMC} to encode both $\bm{z}’_f$ and $\bm{z}’_l$, yielding the corresponding bitstreams $\bm{b}_f$ and $\bm{b}_l$. The arithmetic coding process operates by progressively narrowing an interval within $[0,1)$ based on the predicted symbol probabilities. At the beginning of encoding, the interval is initialized as $I_0=[0,1)$. At each encoding step, the current interval $I_{t-1}=[l_{t-1}, h_{t-1})$, where $l$ and $h$ denote the lower and upper bounds of the interval, respectively, is partitioned into a set of disjoint sub-intervals $\{\tilde{I}_t(x)\}_{x \in \mathbf{A}}$, each corresponding to a distinct symbol from a predefined alphabet $\mathbf{A}$. These sub-intervals are sized in proportion to the conditional probability of each symbol given the preceding context, which can be written as
\begin{align}
	\tilde{I}_t(x) &= \Bigg[ l_{t-1} + (h_{t-1} - l_{t-1}) \times \sum_{m < x} \rho \left(m \mid x_{<t} \right) , \nonumber \\
	&\quad \phantom{=[} l_{t-1} + (h_{t-1} - l_{t-1}) \times \sum_{m \leq x} \rho \left(m \mid x_{<t} \right) \Bigg).
\end{align}
The encoder then selects the sub-interval corresponding to the current symbol and uses it as the updated interval for the next step. After all symbols are processed, the final interval is represented by a binary fraction, yielding the compressed bitstream. Next, we can obtain the two bitstreams $\bm{b}_l$ and $\bm{b}_f$  and transmit them along with the indicator $\bm{S}_j$ to the receiver.

At the decoder, the two bitstreams, $\bm{b}_l$ and $\bm{b}_f$, are first separated using the indicator matrix $\bm{S}_j$. The LM-coded stream $\bm{b}_l$ is decoded iteratively by the LM, which predicts the probability distribution of each character, and by the arithmetic decoder using the predicted distribution. This begins with a predefined placeholder token that initializes the LM.  The LM then predicts the probability distribution of the first symbol, and this distribution, together with the received bitstream $\bm{b}_l$, is provided to the arithmetic decoder to recover the first ASCII character. The recovered character is appended to the context and used by the LM to predict the distribution of the next character. This iterative refinement continues until the entire ASCII sequence has been reconstructed, after which ASCII detokenization is applied to obtain the corresponding quantized values $q_l$. For the FM-coded stream $\bm{b}_f$, parallel decoding can be performed using a unified probability distribution to obtain the corresponding quantized values $q_f$.
These quantized values $q_l$ and $q_f$ are merged back into the quantized CSI feature $\bm{z}^\prime$ according to the indicator matrix $\bm{S}_j$. We then perform dequantization and reconstruct the CSI feature vector, given by
\begin{equation}
		\hat{\bm{z}}=Q_d(\bm{z}^\prime)=\bm{z}^\prime \cdot \Delta + \bm{z}_{min}.
\end{equation}
Finally, the reconstructed CSI matrix $\hat{\bm{H}}_a$ is obtained by feeding $\hat{\bm{z}}$ into the lossy decoder.

\subsubsection{Receiver Complexity Discussion}\label{SCD}
We now briefly discuss the decoding complexity of the proposed TCLNet. Since most of the computational burden arises from the lossless decoding process, while the lossy decoder can be efficiently accelerated by GPUs,
we focus our analysis on the lossless decoding process. As shown in \autoref{graph5}, we define $T_\text{lm}$ and $T_\text{fm}$ as the inference times for the LM and FM, respectively, per token. Additionally, let $Z$ represent the total number of tokens to be decoded, and $c$ denote the proportion of tokens assigned to the FM for parallel decoding. Accordingly, the latency of the lossless decoding process can be expressed as
\begin{equation}
\begin{aligned}
T_{\mathrm{lossless}}
=
\max \Big\{
\underbrace{(1-c)\, Z\, T_{\mathrm{lm}}}_{\text{LM}},
\;
\underbrace{c\, Z\, T_{\mathrm{fm}}}_{\text{FM}}
\Big\}.
\end{aligned}
\end{equation}
where this term captures the maximum latency between the LM and FM decoding processes, since they operate in parallel. We note that the FM is performed in a parallel manner, leading to significantly lower latency compared to the LM, i.e., $T_\text{fm} \ll T_\text{lm}$. Therefore, we can simplify the overall lossless decoding latency to
\begin{equation}
\begin{aligned}
T_{\mathrm{lossless}}
\approx
(1-c)\, Z\, T_{\mathrm{lm}} \leq Z\, T_{\mathrm{lm}}.
\end{aligned}
\end{equation}
Compared to fully autoregressive decoding, which requires $Z\, T_{\mathrm{lm}}$ time, the proposed joint serial and parallel decoding scheme effectively reduces the decoding latency by a factor proportional to $c$. This improves the efficiency of the proposed TCLNet in balancing compression performance and decoding speed. 
\begin{figure*}[t]
	\centering
    \vskip 0.2in
	\centerline{\includegraphics[width=\textwidth]{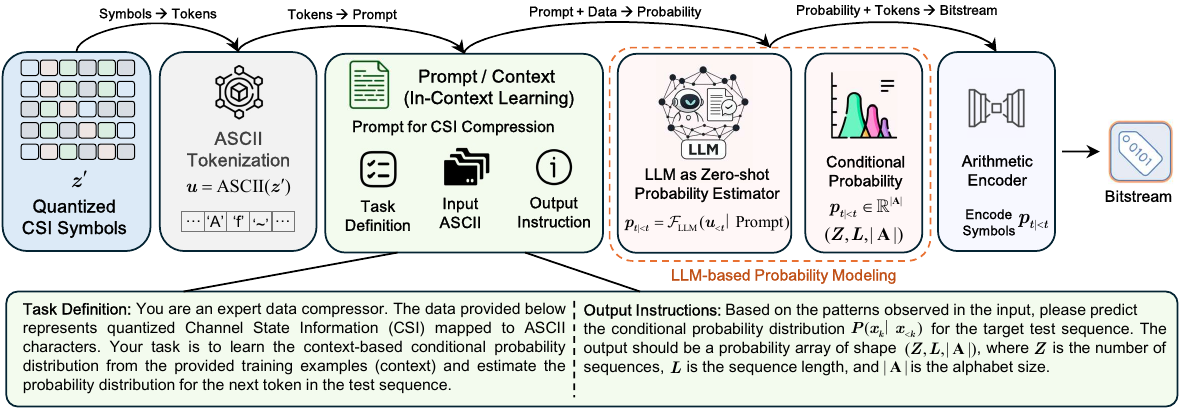}}
	\caption{Illustration of the proposed LLM-based lossless compressor, where ASCII tokenization and in-context learning are employed along with an LLM and an arithmetic encoder.}
	\label{LLM_prompt}
    \vskip -0.2in
\end{figure*}
\begin{figure}[t]
	\begin{center}
    	\vskip 0.2in
		\centerline{\includegraphics[width=\columnwidth]{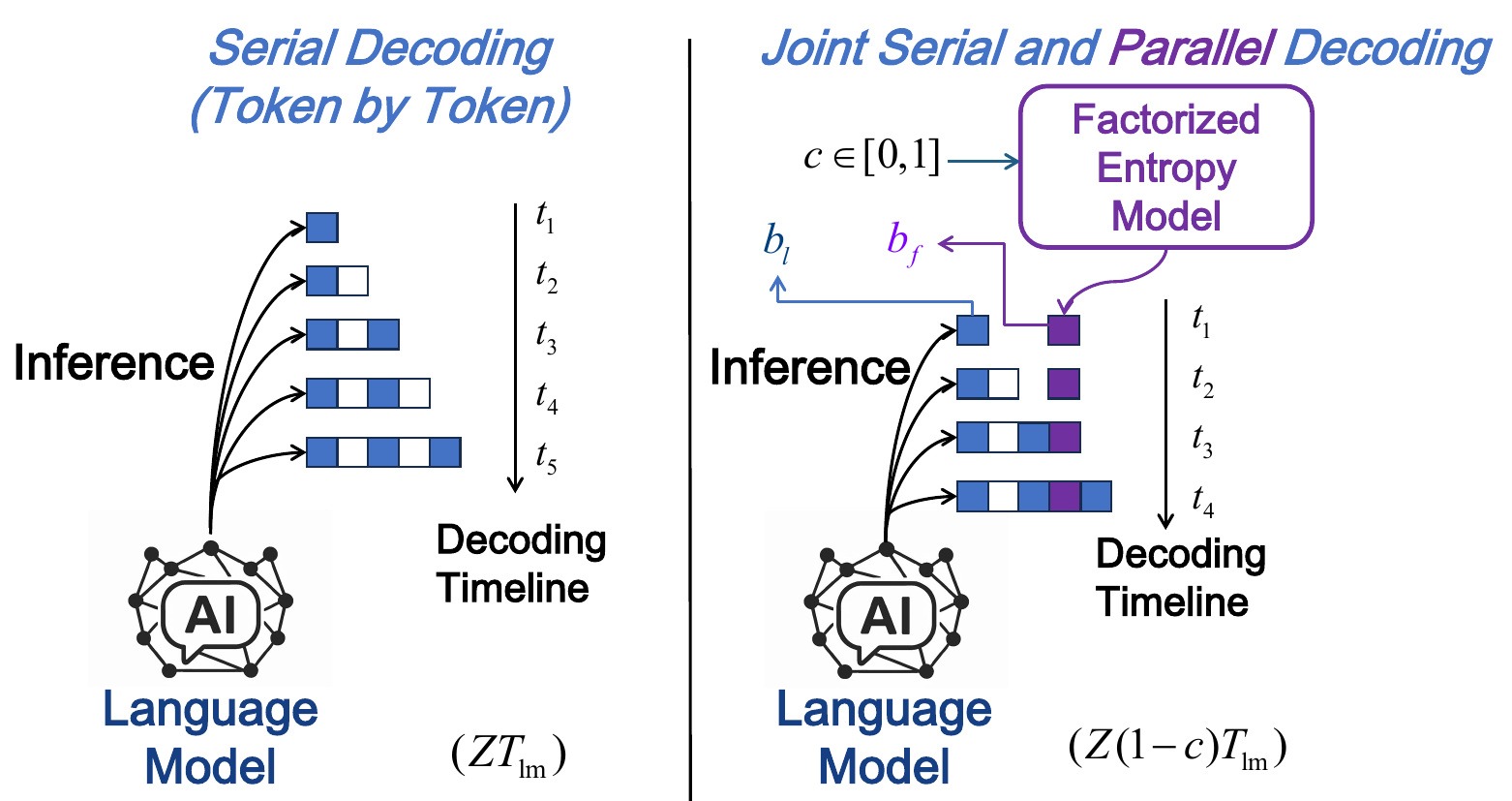}}
		\caption{The proposed joint serial and parallel decoding scheme. The FM parallelly decodes the corresponding tokens, while the LM serially decodes the tokens, with the proportion between them determined by $c$.}
		\label{graph9}
        	\vskip -0.2in
	\end{center}
\end{figure}
\subsection{Large Language Model as Zero-Shot Lossless Compressor}
Despite the high compression efficiency of the LM-based lossless compressor, it requires training on domain-specific CSI datasets. Motivated by the strong in-context learning capabilities of LLMs, we further investigate their potential as zero-shot lossless compressors for CSI feedback, allowing them to model the statistical dependencies of CSI symbols without any parameter updates. The details of the proposed LLM-based lossless compressor are presented in \autoref{LLM_prompt}.

\subsubsection{Workflow and Prompt Engineering}
To align with the textual input space of LLMs, we apply the same ASCII tokenization scheme used in the LM-based compressor, converting the quantized CSI $\bm{z}’$ into discrete character tokens. The probability estimation task is formulated as next-token prediction, where the LLM infers the statistical structure of the target sequence. To make the LLM act as a probability estimator instead of a text generator, we design a simple and structured prompt that explains the task, the input format, and the required output. 




\subsubsection{Probability Extraction and Coding}
Next, we employ the LLM to estimate the conditional probability distributions required for arithmetic coding. Specifically, given input tokenized sequence $\bm{u}$, at each time step $t$, the LLM outputs a conditional probability vector $\bm{p}_t$ based on the preceding context $\bm{u}_{<t}$ and the well-designed prompt, given by
\begin{equation}
    \bm{p}_t = \mathcal{F}_{\text{LLM}}(\bm{u}_{<t} \mid \text{Prompt}),
\end{equation}
where $\mathcal{F}_{\text{LLM}}$ is the forward function of the LLM. The output vector $\bm{p}_t \in \mathbb{R}^{|\mathbf{A}|}$ represents the predicted probability distribution over the vocabulary $\mathbf{A}$ for the next token at position $t$. Finally, these probability vectors $\{\bm{p}_1, \bm{p}_2, \dots, \bm{p}_L\}$ are then sequentially fed into the arithmetic coder to encode the entire sequence into a compact bitstream for transmission.

\section{Experimental Results}
\label{sec:Results}
In this section, we conduct experiments to evaluate the performance of the proposed TCLNet framework. We first introduce the experimental setup. Then, we present and analyze the experimental results in terms of reconstruction accuracy and computational complexity. We also conduct ablation studies to validate the effectiveness of key components in TCLNet.
\subsection{Experimental Setup}
\subsubsection{Datasets}
We evaluate the performance of the proposed TCLNet on both real-world and simulated MIMO channel datasets. For real-world data, we adopt the Argos dataset \cite{Argo}, which was collected at 2.4 GHz across various indoor locations. For simulated data, we utilize channels generated by the COST 2100 model \cite{Csinet} for both indoor and outdoor environments with standard parameter settings.

\subsubsection{Implementation Details}
We implement the proposed TCLNet in PyTorch.
The lossy compression network is trained using the Adam optimizer with a cosine
annealing learning rate schedule, where the initial and minimum learning rates are set to $2\times10^{-3}$ and $5\times10^{-5}$, respectively.
The training process lasts for 500 epochs, including a 20-epoch warm-up. The LM is instantiated as a decoder-only Transformer with an embedding dimension of $256$ and a vocabulary size of 256.
It consists of four Transformer blocks, each equipped with eight attention heads,
and a feed-forward network of dimension 1024.
The LM is trained using Adam with a fixed learning rate of $10^{-4}$ for
8000 iterations with a batch size of 16.
\begin{table*}[!tb]
	\caption{Performance comparison in the Argos indoor scenario. The Bold represents the best results, the underlined represents the second-best.}
	\centering
	\label{Argos_in}
    \begin{threeparttable}
	\begin{tabular}{cccccccccccc}
		\toprule
		\multicolumn{2}{c}{CR} &\multicolumn{2}{c}{1/8}   & \multicolumn{2}{c}{1/16} & \multicolumn{2}{c}{1/32} &\multicolumn{2}{c}{1/64} &\multicolumn{2}{c}{1/128} \\
		\midrule
		Method &Architecture &FLOPs & NMSE & FLOPs & NMSE & FLOPs & NMSE &FLOPs & NMSE &FLOPs & NMSE   \\
		\midrule
		CRNet & CNN  &41.38M & -18.12 &28.92M &-16.01 &22.69M&-11.80 &19.58M &-10.44 &18.02M&-8.36\\
		CLNet & CNN& 36.81M &\underline{-21.80} &24.36M &-16.93&18.13M&-13.39&15.02M &-9.16&13.46M & -6.10 \\
		TransNet &Transformer &822.80M &-19.50 &810.34M &\underline{-17.45}& 804.11M & \textbf{-15.95}& 801.00M &\underline{-13.52}&799.44M&\underline{-11.84}\\
		TCLNet& Trans\&CNN &310.92M&\textbf{-22.64}& 298.46M& \textbf{-17.70}&292.23M&\underline{-15.31}& 289.12M & \textbf{-14.27} &285.18M&\textbf{-11.93}\\
		\midrule
		7 bit-CRNet &CNN &$+ O \left( Q_{7}\right) $ &-18.07& $+ O \left( Q_{7}\right) $ &-15.97 &$+ O \left( Q_{7}\right) $ &-11.78&$+ O \left( Q_{7}\right) $ &-10.28&$+ O \left( Q_{7}\right) $ &-8.22\\
		7 bit-CLNet &CNN &$+ O \left( Q_{7}\right) $ &\underline{-21.52}& $+ O \left( Q_{7}\right) $ &-16.93 &$+ O \left( Q_{7}\right) $ &-13.36&$+ O \left( Q_{7}\right) $ &-9.19&$+ O \left( Q_{7}\right) $ &-6.09\\
		7 bit-TransNet&Transformer &$+ O \left( Q_{7}\right) $ &-19.37& $+ O \left( Q_{7}\right) $ &\underline{-17.42} &$+ O \left( Q_{7}\right) $ &\textbf{-15.80}&$+ O \left( Q_{7}\right) $ &\underline{-13.63}&$+ O \left( Q_{7}\right) $ &\underline{-11.82} \\
		7 bit-TCLNet&Trans\&CNN &$+ O \left( Q_{7}\right) $ &\textbf{-22.41}& $+ O \left( Q_{7}\right) $ &\textbf{-17.64} &$+ O \left( Q_{7}\right) $ &\underline{-15.26}&$+ O \left( Q_{7}\right) $ &\textbf{-14.22}&$+ O \left( Q_{7}\right) $ &\textbf{-11.91} \\
		\midrule
		6 bit-CRNet&CNN&$+ O \left( Q_{6}\right) $ &-17.91&$+ O \left( Q_{6}\right) $ &-15.87&$+ O \left( Q_{6}\right) $ &-11.74&$+ O \left( Q_{6}\right) $ &-10.24&$+ O \left( Q_{6}\right) $ &-8.18\\
		6 bit-CLNet&CNN &$+ O \left( Q_{6}\right) $ &\underline{-20.76}&$+ O \left( Q_{6}\right) $ &-16.73&$+ O \left( Q_{6}\right) $ &-13.28&$+ O \left( Q_{6}\right) $ &-9.16&$+ O \left( Q_{6}\right) $ &-6.08\\
		6 bit-TransNet&Transformer &$+ O \left( Q_{6}\right) $ &-19.06&$+ O \left( Q_{6}\right) $ &\underline{-17.21}&$+ O \left( Q_{6}\right) $ &\textbf{-15.67}&$+ O \left( Q_{6}\right) $ &\underline{-13.58}&$+ O \left( Q_{6}\right) $ &\underline{-11.78}\\
		6 bit-TCLNet&Trans\&CNN &$+ O \left( Q_{6}\right) $ &\textbf{-21.80}&$+ O \left( Q_{6}\right) $ &\textbf{-17.47}&$+ O \left( Q_{6}\right) $ &\underline{-15.18}&$+ O \left( Q_{6}\right) $ &\textbf{-14.15}&$+ O \left( Q_{6}\right) $ &\textbf{-11.84}\\
		\midrule
		5 bit-CRNet &CNN&$+ O \left( Q_{5}\right) $ &-17.34&$+ O \left( Q_{5}\right) $ &-15.45&$+ O \left( Q_{5}\right) $ &-11.59&$+ O \left( Q_{5}\right) $ &-10.09&$+ O \left( Q_{5}\right) $ &-8.02\\
		5 bit-CLNet&CNN &$+ O \left( Q_{5}\right) $ &\underline{-18.61}&$+ O \left( Q_{5}\right) $ &-15.99&$+ O \left( Q_{5}\right) $ &-12.97&$+ O \left( Q_{5}\right) $ &-9.02&$+ O \left( Q_{5}\right) $ &-6.02\\
		5 bit-TransNet &Transformer &$+ O \left( Q_{5}\right) $ &-17.93&$+ O \left( Q_{5}\right) $ &\underline{-16.17}&$+ O \left( Q_{5}\right) $ &\textbf{-15.61}&$+ O \left( Q_{5}\right) $ &\underline{-13.38}&$+ O \left( Q_{5}\right) $ &\textbf{-11.62}\\
		5 bit-TCLNet&Trans\&CNN &$+ O \left( Q_{5}\right) $ &\textbf{-20.06}&$+ O \left( Q_{5}\right) $ &\textbf{-16.85}&$+ O \left( Q_{5}\right) $ &\underline{-14.87}&$+ O \left( Q_{5}\right) $ &\textbf{-13.87}&$+ O \left( Q_{5}\right) $ &\underline{-11.58}\\
		\bottomrule
        
	\end{tabular}
    \begin{tablenotes}[flushleft]
		\footnotesize
		\item \textit{Notes:} To illustrate the effectiveness of the lossy compressor, we do not employ   the proposed lossless compressor here.
	\end{tablenotes}
	\end{threeparttable}
\end{table*}

\subsubsection{Baselines}
We compare the proposed TCLNet with several representative deep learning-based CSI feedback methods with varying levels of complexity. For low-complexity methods, we compare with three CNN-based networks: CsiNet \cite{Csinet}, CRNet \cite{CRNet}, and CLNet \cite{CLNet}. For higher-complexity methods, we provide comparisons with TransNet \cite{Transnet} and CsiNet+ \cite{Csinet+}.

\subsubsection{Evaluation Metrics}
We evaluate CSI feedback performance from three perspectives: reconstruction accuracy, computational complexity, and transmission efficiency. Reconstruction accuracy is measured using the normalized mean square error (NMSE), given by
\begin{align}
\text{NMSE} = \mathbb{E} \left[ \frac{\lVert \bm{H}_a - \hat{\bm{H}}_a \rVert_2^2}{\lVert \bm{H}_a \rVert_2^2} \right].
\end{align}
Computational complexity is quantified by the number of floating-point operations (FLOPs) required during inference. Transmission efficiency is assessed by the number of bits transmitted per CSI value. Let $N$ denote the number of CSI elements \cite{Csinet} and $B$ the total transmitted bits, the bit rate per CSI value can be given by
\begin{align}
\text{Bit Rate} = \frac{B}{N}.
\end{align}

\subsection{Results on Real-World Datasets}
\begin{table*}[!tb] 	
	\caption{Encoding results of TCLNet Versus conventional fixed-length coding on the Argos dataset with 7-bit uniform quantization.}
	\centering
	\label{tab_codec}
	\begin{tabular}{cccccccc}
		\toprule
		\multirow{3}{*}{Method} & \multirow{3}{*}{Codec} & \multirow{3}{*}{Decoding Time} & \multicolumn{5}{c}{CR / Bit Rate} \\ 
		\cmidrule(lr){4-8}
		& & & 1/8 & 1/16 & 1/32 & 1/64 & 1/128 \\
		\midrule
		CRNet   &\multirow{3}{*}{Fixed-Length Codec}        &\multirow{3}{*}{ $\approx T_\text{AD} \times Z$} &\multirow{3}{*}{0.8750} &\multirow{3}{*}{0.4375} &\multirow{3}{*}{0.2188} &\multirow{3}{*}{0.1094} &\multirow{3}{*}{0.0546}\\
		CLNet  &      & & &&& \\
		TransNet & &  & &&&&\\
		\noalign{\vskip 3pt}  
		\hdashline
		\noalign{\vskip 3pt}  
		TCLNet ($c=1$)   & \multirow{6}{*}{Entropy Codec} & $T_\text{AD}\times Z+T_\text{fm}$                            & 0.7732 & 0.3985 & 0.2000 & 0.1038 & 0.0534 \\
		TCLNet ($c=0.8$) & & $T_\text{AD}\times Z+T_\text{fm}\times0.8 + T_\text{lm}\times0.2$ & 0.7280 & 0.3765 & 0.1860 & 0.0976 & 0.0497 \\
		TCLNet ($c=0.6$) & & $T_\text{AD}\times Z+T_\text{fm}\times0.6 + T_\text{lm}\times0.4$ & 0.6820 & 0.3560 & 0.1733 & 0.0930 & 0.0471 \\
		TCLNet ($c=0.4$) &  & $T_\text{AD}\times Z+T_\text{fm}\times0.4 + T_\text{lm}\times0.6$ & 0.6380 & 0.3360 & 0.1608 & 0.0885 & 0.0452 \\
		TCLNet ($c=0.2$) & & $T_\text{AD}\times Z+T_\text{fm}\times0.2 + T_\text{lm}\times0.8$ & 0.5940 & 0.3150 & 0.1485 & 0.0840 & 0.0432 \\
		TCLNet ($c=0$)   &  & $T_\text{AD}\times Z+T_\text{lm}$                           & \textbf{0.5470} & \textbf{0.2950} & \textbf{0.1358} & \textbf{0.0799} & \textbf{0.0417} \\
		\noalign{\vskip 3pt}  
		\hdashline
		\noalign{\vskip 3pt}  
		
		\multicolumn{3}{c}{Estimated Entropy (Lower Bound, $c=0$)}                           & 0.5459 & 0.2939 & 0.1332 & 0.0790 & 0.0416 \\
		\multicolumn{3}{c}{Compression Efficiency ($c=0$)}                         & \textbf{99.79\%} & \textbf{99.32\%} & \textbf{98.09\%} & \textbf{98.87\%} & \textbf{99.76\%} \\

		\bottomrule
	\end{tabular}
\end{table*}

In Table \ref{Argos_in}, we compare the performance of TCLNet with baseline methods in terms of NMSE and FLOPs on the real-world Argos dataset, where the compression ratio (CR) is set to 1/8, 1/16, 1/32, 1/64, and 1/128. The input CSI matrix has a dimension of $96 \times 52$, which is more representative of practical large-scale MIMO systems compared to the commonly used $32 \times 32$ representation in prior works such as \cite{Transnet}. From this table, we can observe that the proposed TCLNet consistently outperforms CNN-based methods across all CR settings, and also outperforms the pure Transformer-based TransNet in most scenarios with significantly lower computational complexity. Specifically, at the CR of 1/8, the proposed TCLNet improves NMSE by 4dB at most compared to CNN-based methods. Moreover, the proposed TCLNet also achieves gains of 1-2 dB over TransNet which uses a pure Transformer architecture, while requiring only about 
1/3 of its FLOPs. In the lower CR regimes, e.g., CR=1/64 and 1/128, the proposed TCLNet still achieves about 0.5 dB improvement over TransNet with much lower complexity. This is because the hybrid Trans\&Conv architecture effectively captures both global and local features of CSI, while avoiding the high computational overhead associated with deep Transformer models. These results demonstrate the effectiveness of the proposed TCLNet and its suitability for deployment in practical large-scale MIMO systems.

In Table \ref{tab_codec}, we compare the lossless coding performance of the proposed TCLNet with conventional fixed-length coding methods on the Argos dataset under 7-bit quantization, where CR varies from 1/8 to 1/128, and $c$ is set to 0, 0.2, 0.4, 0.6, 0.8, and 1. We report the decoding time complexity and the achieved bit rate for each method. The decoding time is expressed in terms of the arithmetic decoding time $T_\text{AD}$, $T_\text{fm}$, and $T_\text{lm}$, representing the arithmetic decoding time per token, the FM inference time, and the LM inference time, respectively. From this table, we can observe that the proposed TCLNet with entropy coding consistently outperforms fixed-length coding methods across all CR and $c$ settings in terms of bit rate. When $c$ approaches 1, indicating that the FM is primarily used for encoding, the proposed TCLNet still achieves a significant reduction in bit rate compared to fixed-length coding. When the receiver has sufficient computational resources, choosing a smaller $c$ allows the LM to participate more heavily in probability modeling, thereby further improving compression efficiency. In particular, when $c=0$, the proposed TCLNet achieves the lowest bit rates of 0.5470 with a CR of 1/8, which is very close to the estimated entropy lower bound. These results demonstrate that the proposed lossless compressor can effectively exploit the contextual dependencies in CSI data to achieve superior compression performance, and the proposed complexity-controllable scheme allows for flexible adaptation to varying computational resource constraints at the receiver.

\begin{figure*}[!tb]
    \centering
    \vskip -0.2in
    \subfloat[\footnotesize 7-bit Quantization.]{%
        \includegraphics[width=0.333\textwidth]{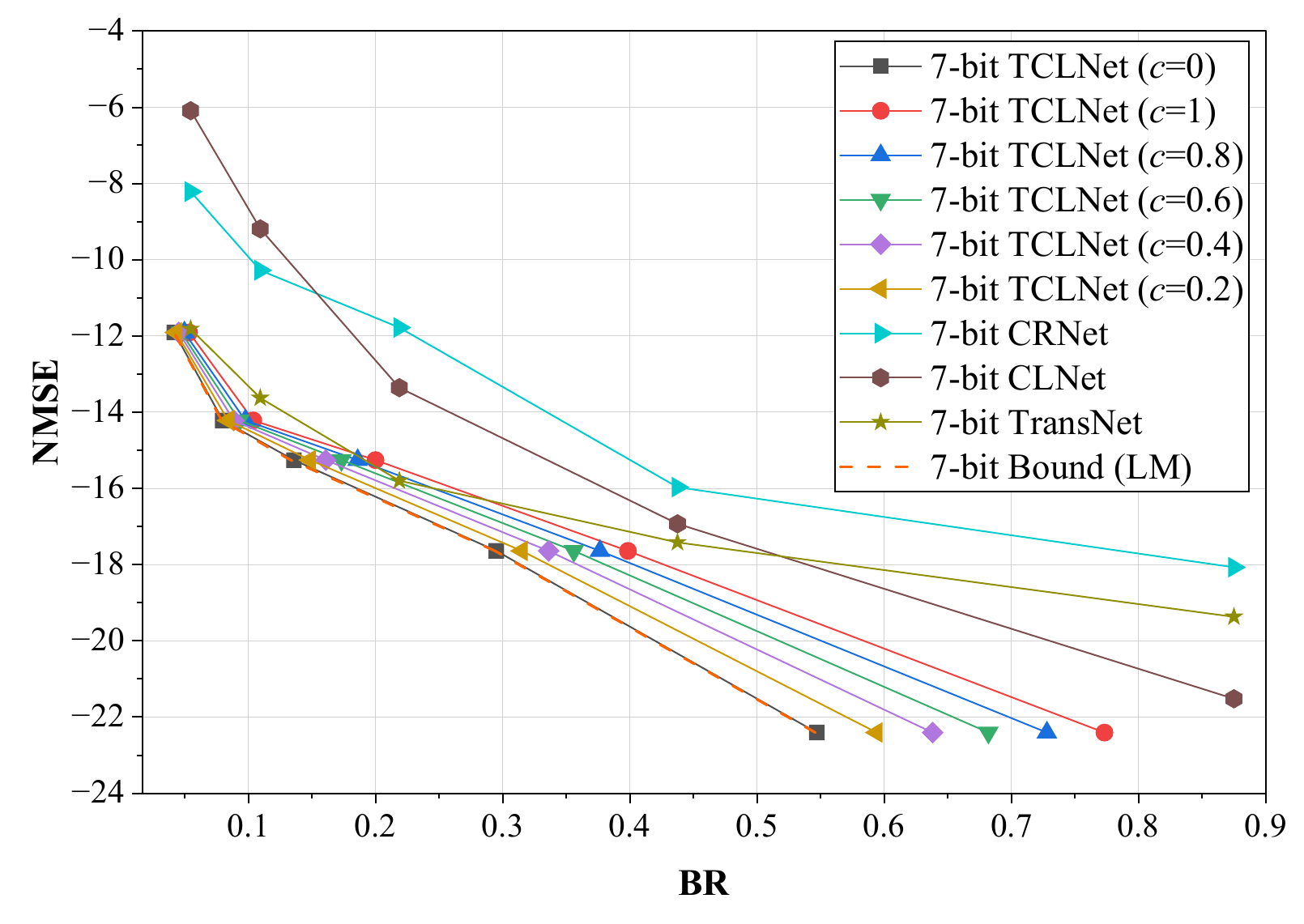}%
        \label{7-a}}
    \hfill
    \subfloat[\footnotesize 6-bit Quantization.]{%
        \includegraphics[width=0.333\textwidth]{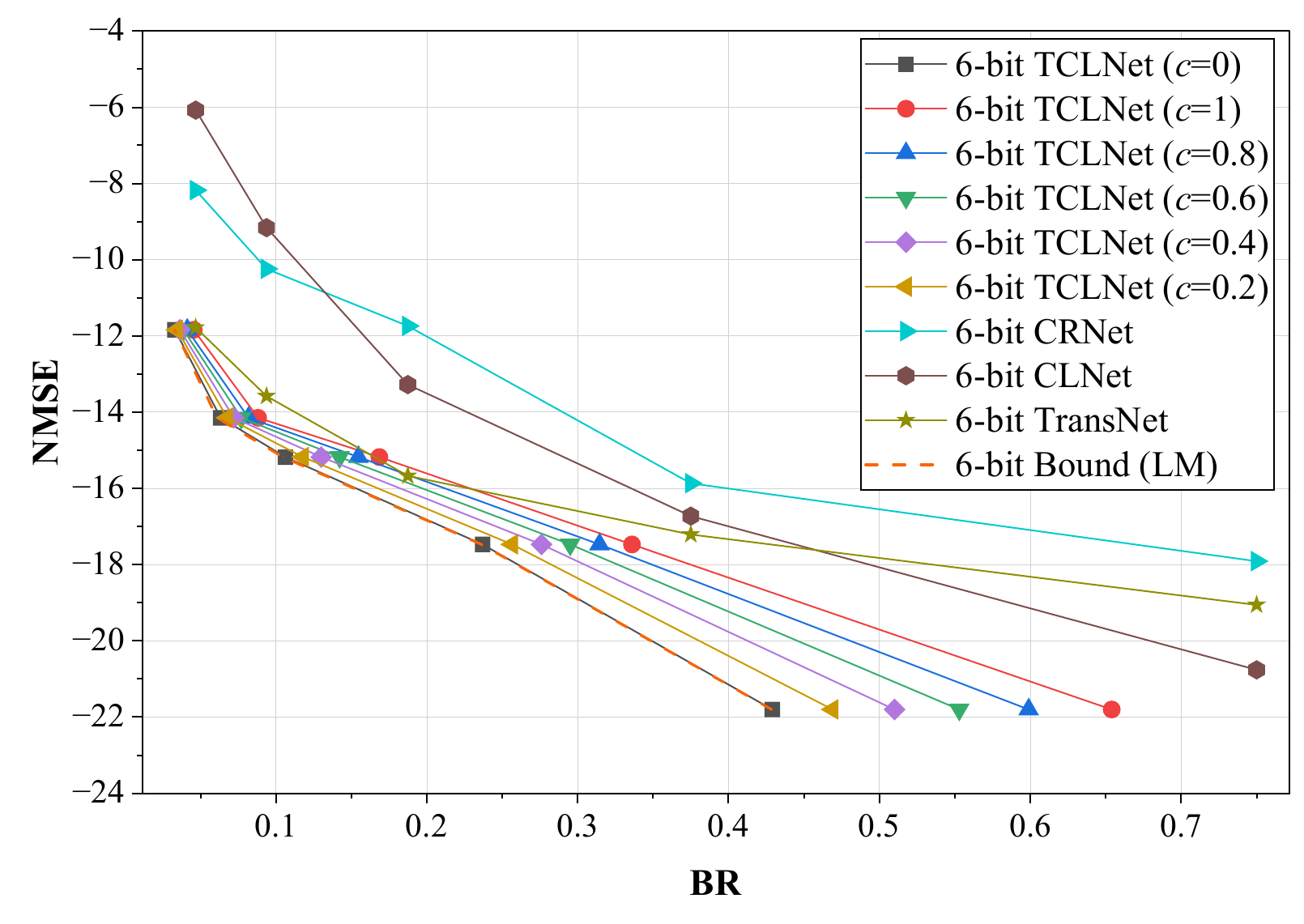}%
        \label{7-b}}
    \hfill
    \subfloat[\footnotesize 5-bit Quantization.]{%
        \includegraphics[width=0.333\textwidth]{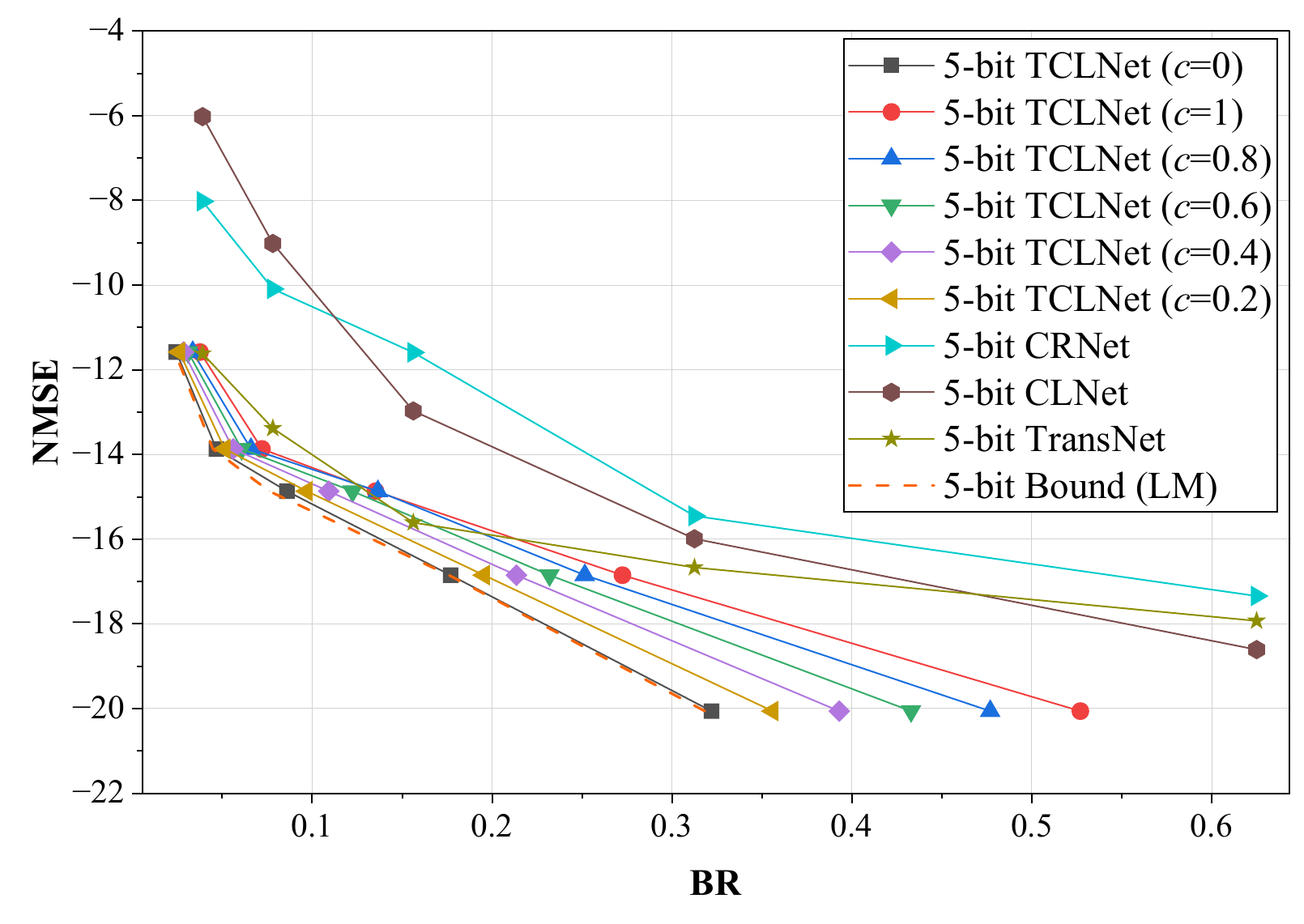}%
        \label{7-c}}
    \caption{RDC trade-off curves on the Argos dataset with different bit quantizations. 
    A higher value of $c$ indicates a lower decoding complexity.}
    \label{RDC-tradeoff}
    	\vskip -0.2in
\end{figure*}

In \autoref{RDC-tradeoff}, we further present the RDC trade-off curves of TCLNet under different quantization settings on the Argos dataset, where $c$ varies from 0 to 1 to represent different decoding complexities. We also provide the performance of baseline methods and the theoretical compression bound estimated by the trained LM for comparison. From the figures, we observe that TCLNet consistently achieves a favorable RDC trade-off across different quantization settings and outperforms baseline methods such as CRNet, CLNet, and TransNet at most bit rates. Moreover, TCLNet maintains performance much closer to the theoretical entropy bound, particularly when $c$ is small, demonstrating the effectiveness of the LM in exploiting contextual dependencies within CSI. Another important observation is that the choice of $c$ should be adapted based on the quantization level. Specifically, for high-bit quantization, using a smaller $c$ yields substantial compression gains, as the LM can effectively model fine-grained symbol correlations. For low-bit quantization, the performance gap between $c=0$ and $c=1$ narrows, suggesting that higher $c$ (lower complexity) is sufficient and preferable for reducing decoding overhead. These results further validate the flexibility and effectiveness of the proposed TCLNet framework in making a good RDC trade-off.

\subsection{Results on Simulated Datasets}
We further evaluate the proposed TCLNet on the widely-used simulated COST2100 dataset \cite{Csinet}, with the input CSI matrix dimension set to $32 \times 32$. As shown in Table \ref{flops_comparison}, we first present the indoor NMSE performance and computational complexity of TCLNet compared to baseline methods under various CR ranging from 1/4 to 1/64. It is observed that while lightweight CNN-based methods such as CLNet and CRNet maintain the lowest complexity, they often suffer from limited reconstruction performance. Among the high-performance models, the proposed TCLNet achieves competitive NMSE results compared to TransNet and CsiNet+. Specifically, at CR=1/4, TCLNet attains an NMSE of -29.29 dB, outperforming TransNet's -29.24 dB and significantly surpassing CsiNet+'s -27.37 dB. This trend continues across other CR settings. Notably, at CR=1/64, TCLNet achieves the best NMSE of -7.81 dB, indicating its robustness in high compression scenarios. These results highlight the effectiveness of the TCLNet architecture in balancing reconstruction accuracy and model complexity.

\begin{table}[t]
\caption{NMSE (dB) and computational complexity (M FLOPs) comparison on the COST2100 dataset.}
\label{flops_comparison}
\centering
\resizebox{\columnwidth}{!}{
\begin{tabular}{cc ccccc}
    \toprule
    \multirow{2}{*}{\textbf{Method}} & \multirow{2}{*}{\textbf{Metric}} & \multicolumn{5}{c}{\textbf{Compression Rates (CRs)}} \\
    \cmidrule(lr){3-7}
     & & \textbf{1/4} & \textbf{1/8} & \textbf{1/16} & \textbf{1/32} & \textbf{1/64} \\
    \midrule
    \multirow{2}{*}{CLNet \cite{CLNet}} & NMSE & -29.16 & -15.60 & -11.15 & -8.95 & \underline{-6.34} \\
     & FLOPs & 4.05M & 3.01M & 2.48M & 2.22M & 2.09M \\
    \midrule
    \multirow{2}{*}{CRNet \cite{CRNet}} & NMSE & -26.99 & -16.01 & -11.35 & -8.93 & -6.49 \\
     & FLOPs & 5.12M & 4.07M & 3.55M & 3.29M & 3.16M \\
    \midrule
    \multirow{2}{*}{CsiNet+ \cite{Csinet+}} & NMSE & -27.37 & -18.29 & -14.14 & \textbf{-10.43} & -- \\
     & FLOPs & 24.57M & 23.52M & 23.00M & 22.74M & -- \\
    \midrule
    \multirow{2}{*}{TransNet \cite{Transnet}} & NMSE & \underline{-29.24} & \textbf{-22.84} & \textbf{-14.96} & \underline{-10.35} & -6.08 \\
     & FLOPs & 35.72M & 34.70M & 34.14M & 33.88M & 33.75M \\
    \midrule
    \multirow{2}{*}{TCLNet} & NMSE & \textbf{-29.29} & \underline{-21.67} & \underline{-14.77} & -10.19 & \textbf{-7.81} \\
     & FLOPs & 31.47M & 30.42M & 29.89M & 29.64M & 29.50M \\
    \bottomrule
\end{tabular}
}
\end{table}

\begin{figure}[t!]
	\begin{center}
    	\vskip 0.2in
		\centerline{\includegraphics[width=\columnwidth]{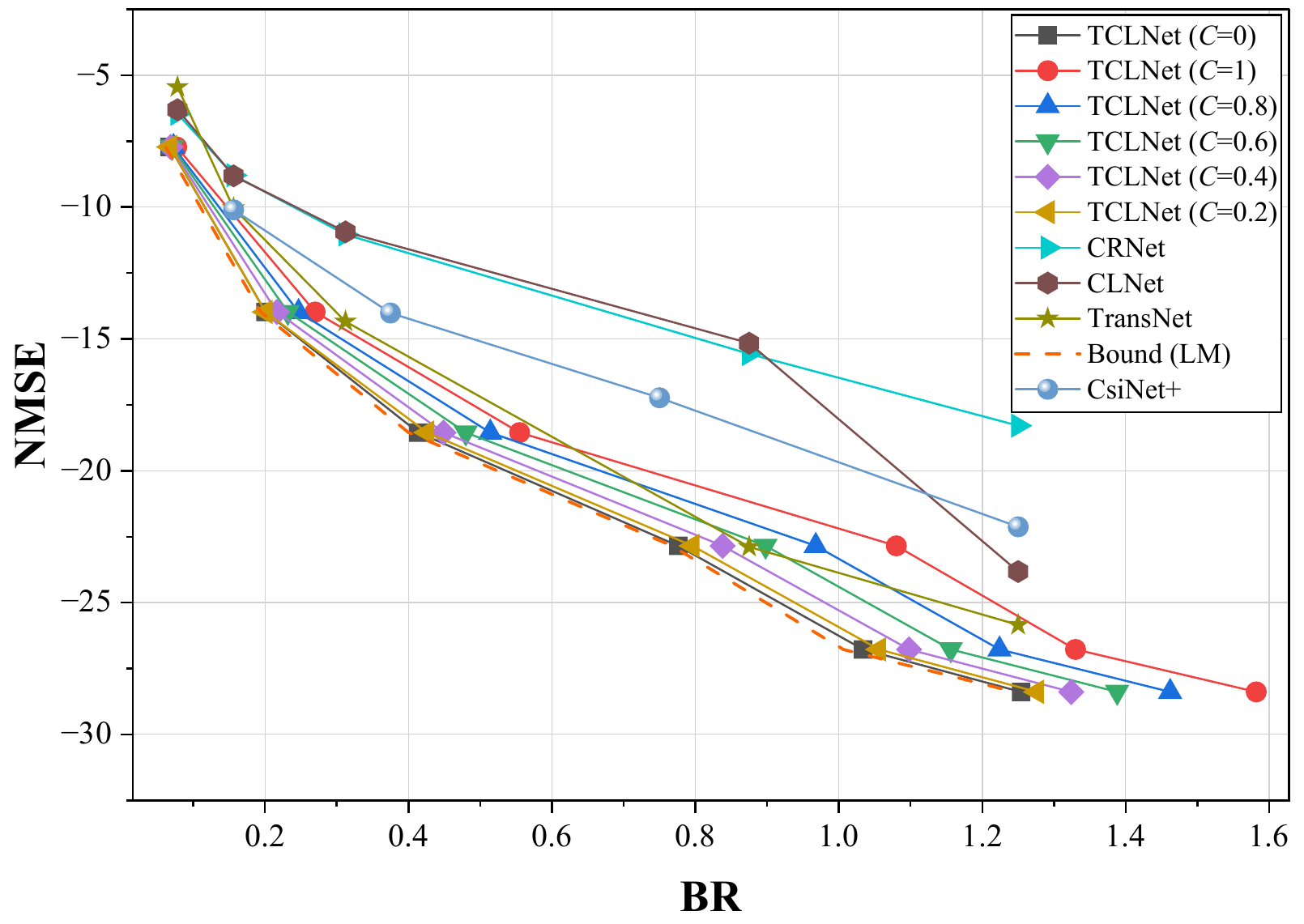}}
		\caption{Bit Rate-NMSE trade-off comparison on the COST2100 indoor dataset. TCLNet achieves superior reconstruction accuracy, particularly at higher bit rates.}
		\label{graph10}
        	\vskip -0.2in
	\end{center}
\end{figure}

\begin{figure}[t!]
	\begin{center}
    	\vskip 0.2in
		\centerline{\includegraphics[width=\columnwidth]{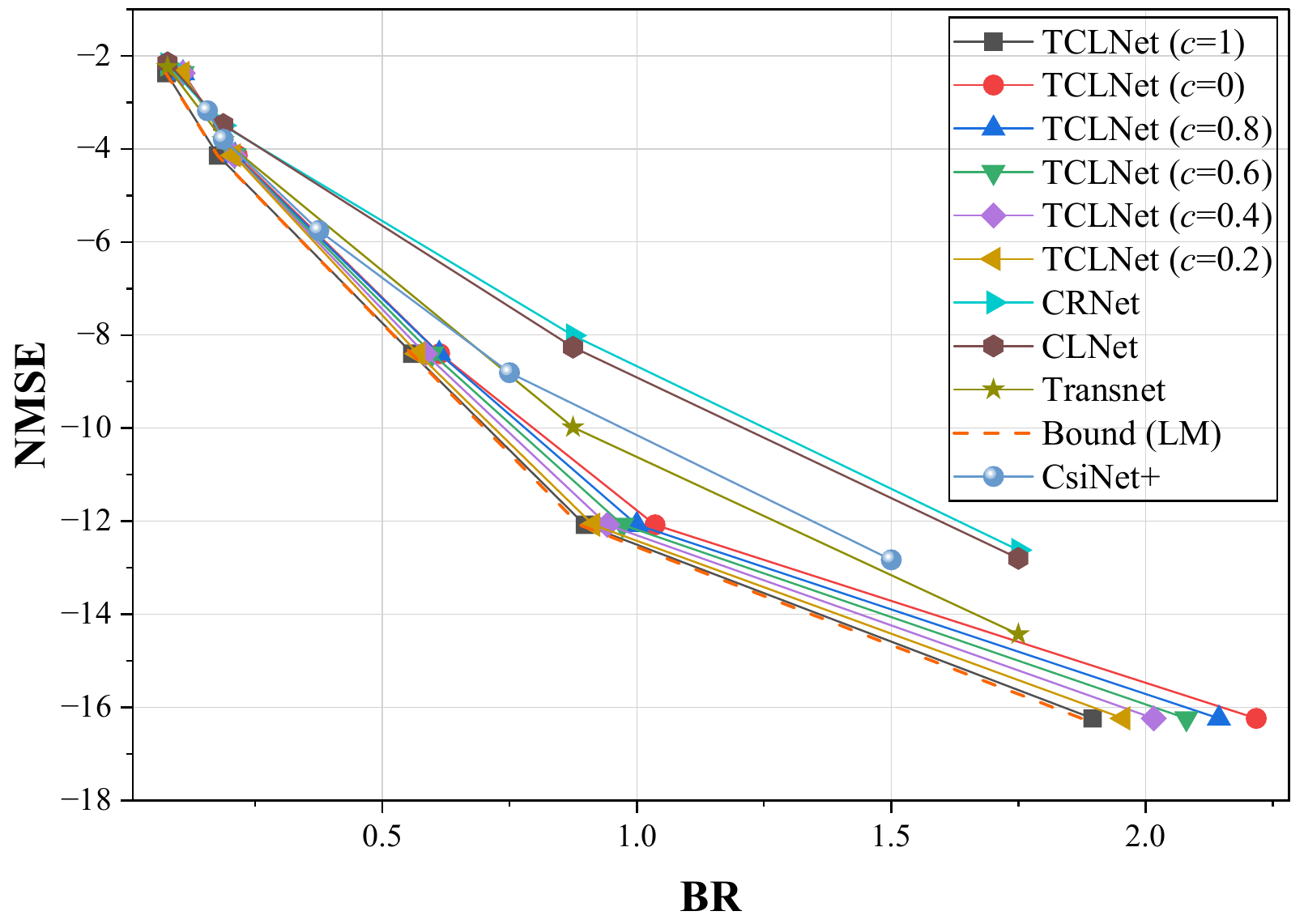}}
		\caption{Bit Rate-NMSE trade-off comparison on the COST2100 outdoor dataset. TCLNet demonstrates robust performance even under low bit-rate constraints.}
		\label{graph11}
        	\vskip -0.2in
	\end{center}
\end{figure}

In \autoref{graph10} and \autoref{graph11}, we present the BR versus NMSE trade-off curves of TCLNet and baseline methods on the COST2100 indoor and outdoor datasets, respectively. It can be observed that at low bit rates, TCLNet outperforms CsiNet+ and TransNet, and significantly surpasses CLNet and CRNet. As the bit rate increases, the performance gain of TCLNet becomes more pronounced. This can be attributed to the enhanced representation capability of TCLNet and the effective lossless compression achieved by the proposed entropy coder. Besides, by setting different values of $c$, we can observe how closely TCLNet approaches the performance upper bound when more symbols are decoded using the LM, e.g., a lower $c$ value. While higher $c$ values lead to reduced complexity, however, the proposed TCLNet still maintains competitive performance. Furthermore, similar trends are observed in the outdoor scenarios, where TCLNet consistently outperforms baseline methods across various bit rates and reaches close to the theoretical performance limit. These results demonstrate the robustness and effectiveness of the proposed TCLNet architecture and its ability to adaptively balance compression efficiency and computational complexity across different scenarios.

\subsection{	Robustness to Channel Estimation Errors}
In \autoref{Robust}, we evaluate the robustness of different methods against downlink channel estimation errors on the Argos dataset. The estimated downlink CSI at the user is modeled as being corrupted by additive white Gaussian noise (AWGN) with an input SNR of 5 dB, and the CR is set to $1/64$. From the results, we can observe that all methods exhibit a degradation in reconstruction accuracy when noisy CSI is used as input, and the proposed TCLNet still achieves the lowest NMSE among all competitors, outperforming both CNN-based and Transformer-based baselines. These results demonstrate the strong robustness of TCLNet against channel estimation errors.
\begin{table}[t!]
	\caption{Robustness Against Channel Estimation Error}
	\label{Robust}
	\centering
	\renewcommand{\arraystretch}{1.2} 
	\setlength{\tabcolsep}{8pt}   
		\begin{tabularx}{0.8\linewidth}{ccccc}
			\toprule
			Method & FLOPs & \makecell{NMSE/dB \\(Perfect CSI)}& \makecell{NMSE/dB \\(Noisy CSI)}\\
			\midrule
			CRNet & 19.579M & -10.44 & -10.18 \\
			CLNet & 15.016M & -9.159 & -8.863 \\
			TransNet & 800.996M &\underline{-13.52} & \underline{-13.45}\\
			\textbf{TCLNet} & {289.117M} &\textbf{-14.27} &\textbf{ -13.67} \\
			\bottomrule
		\end{tabularx}
\end{table}
\subsection{Impact of Window Size in TCLNet}
\begin{figure}[t!]
	\begin{center}
    	\vskip 0.2in
		\centerline{\includegraphics[width=\columnwidth]{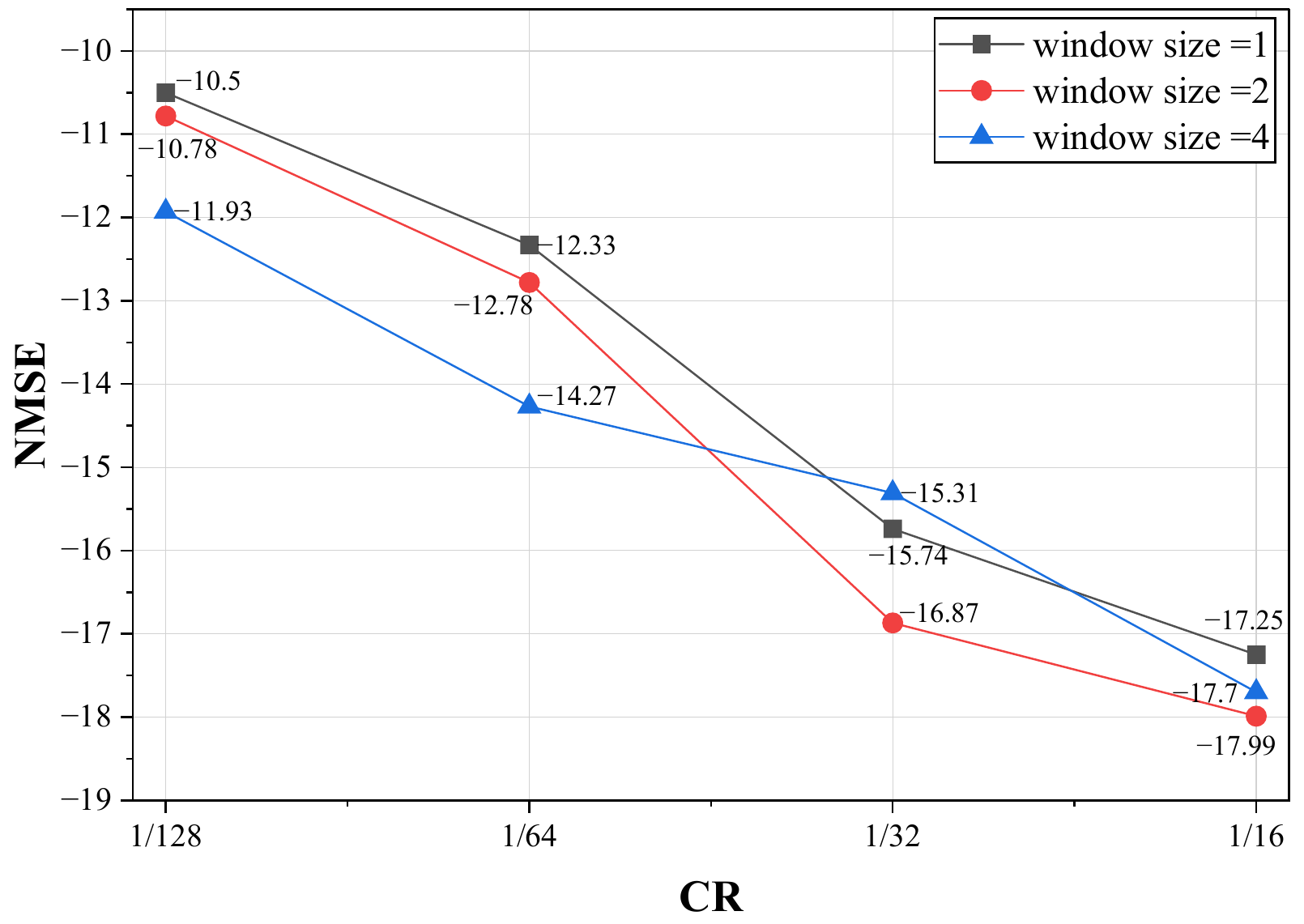}}
		\caption{NMSE Comparison of TCLNet under different window sizes, where CR ranges from 1/16 to 1/128.}
		\label{window}
        \vskip -0.2in
	\end{center}
\end{figure}

In \autoref{window}, we investigate the effect of different window-size settings in the transformer module of the proposed TCLNet on the Argos dataset, where the CR varies from 1/16 to 1/128. We can observe that the window size significantly influences the reconstruction performance of TCLNet. Specifically, a window size of 4 consistently yields the best NMSE across most compression ratios, as it effectively captures the spatial dependencies within the CSI matrix through self-attention mechanisms. However, it also incurs the risk of overfitting, as seen when CR$=1/32$, where a window size of 4 performs slightly worse than smaller window sizes. When the window size is reduced to 2, the model still benefits from self-attention, but to a lesser extent, leading to moderate performance. Conversely, a window size of 1 effectively reduces the self-attention mechanism to a convolutional operation, which may not fully exploit the spatial correlations in the CSI data, resulting in the poorest performance among the three configurations.

\subsection{Potential of LLMs as Lossless CSI Compressors}\label{LLM_C}
\begin{table*}[tb]
	\caption{Comparison of compression performance using prompt-enabled LLM as a lossless compressor.}
	\begin{center}
		\label{LLM_as_compressor}
		\begin{tabular}{ccccccc}
			\toprule
			Model &Fixed-Length Scheme & FM & LM &ChatGPT-5 & ChatGPT-4o  \\
			\midrule
			Model State& --& \makecell{Trained with \\ CSI Symbols }& \makecell{Trained with \\ CSI Symbols (ASCII)} &\makecell{Pretrained \\with Text} &\makecell{Pretrained \\with Text} \\
			\midrule
			Method&Fixed Bit Rate &Training&Training&Prompt &Prompt \\
			\midrule
			Parameters&--&$\approx$ 20K&$\approx$ 12.5M&$>$ 1T&$\approx$ 200B\\
			\midrule
			Bit Rate&0.0546&0.0534&0.0417&0.0513&0.0515\\
			\makecell{Estimated Entropy} &0.0546&0.0520&0.0416&0.0498&0.0501\\
			\bottomrule
		\end{tabular}
	\end{center}
\end{table*}

In Table \ref{LLM_as_compressor}, we finally explore the potential of LLMs as lossless compressors for CSI feedback. We compare the compression performance of two pretrained LLMs, ChatGPT-5 and ChatGPT-4o, with that of traditional fixed-length coding schemes, FM-based coding, and a Transformer-decoder trained on CSI symbols using ASCII tokenization. We use the Argos dataset for evaluation, where a lossy compressor with a compression ratio of 128 produces symbol sequences of length 78. From the results, we observe that ChatGPT-5 and ChatGPT-4o achieve bit rates of 0.0513 and 0.0515, respectively, outperforming the fixed-length coding scheme and FM-based coding, but slightly underperforming the Transformer-decoder trained on CSI symbols. These results highlight that although general-purpose LLMs are not specifically trained on CSI data, their strong contextual modeling capabilities already enable competitive compression performance, suggesting their promising potential as universal lossless compressors for future CSI feedback systems.

\section{Conclusion}
\label{sec:conclusion}

In this paper, we have proposed a novel CSI feedback network, TCLNet. By leveraging a hybrid transformer-CNN architecture, TCLNet achieves efficient CSI compression while significantly reducing computational complexity compared to prior transformer-based models. In addition, we have integrated both LM- and FM-based lossless coding schemes into TCLNet, enabling it to achieve superior RDC trade-offs for bit-level CSI compression. We have also explored the potential of LLMs as universal lossless compressors for CSI feedback. We have conducted extensive experiments on multiple datasets, demonstrating the superiority of the proposed TCLNet over existing data-driven approaches for CSI feedback and fixed-length coding methods. The results also highlight the promising potential of LLMs in this novel application domain. Future work may investigate several directions to further enhance system performance. First, parameter-efficient fine-tuning techniques such as LoRA can be employed to specialize LLMs for CSI probability modeling without incurring significant training overhead. Second, advanced attention mechanisms, such as cross-attention, can be explored to further improve the modeling of spatial dependencies in CSI data. 

\bibliography{IEEEabrv,refe.bib}

\bibliographystyle{IEEEtran}

\end{document}